\documentclass[twocolumn,superscriptaddress]{revtex4}
\usepackage[english]{babel}
\usepackage{amsmath, amsfonts, amssymb, bm, graphicx, color}

\newcommand{\IN}{}
\newcommand{\FC}{}

\newcommand{\apdx}{Appendix}
\renewcommand{\vec}{\boldsymbol}
\newcommand{\DKL}{\ensuremath{{D_{\rm KL}}}}
\newcommand{\Hell}{\ensuremath{{D_{\rm H}}}}
\newcommand{\Bhatt}{\ensuremath{{\rm BC}}}
\newcommand{\Dir}[2]{\ensuremath{{\rm Dir}(\vec{#1}\vert {#2})}}
\newcommand{\Mult}[2]{\ensuremath{{\rm Mult}(\vec{#1}\vert\vec{#2})}}

\newcommand{\lastequal}{These authors contributed equally.}
\newcommand{\corr}{To whom correspondence should be addressed.}

\begin{document}

\newcommand{\deftitle}{{Bayesian estimation of the Kullback-Leibler divergence for categorical sytems\\
using mixtures of Dirichlet priors}}

\title{\deftitle}

\author{Francesco Camaglia}
\affiliation{Laboratoire de physique de l'\'Ecole normale sup\'erieure,
  CNRS, PSL University, Sorbonne Universit\'e and Universit\'e de
  Paris, 75005 Paris, France}
\author{Ilya Nemenman}
\affiliation{
Department of Physics, Department of Biology, and
 Initiative for Theory and Modeling of Living Systems, Emory University, Atlanta, Georgia, USA
}
\author{Thierry Mora}
\thanks{\corr}
\affiliation{Laboratoire de physique de l'\'Ecole normale sup\'erieure,
  CNRS, PSL University, Sorbonne Universit\'e and Universit\'e de
  Paris, 75005 Paris, France}
\affiliation{\lastequal}
\author{Aleksandra M. Walczak}
\thanks{\corr}
\affiliation{Laboratoire de physique de l'\'Ecole normale sup\'erieure,
  CNRS, PSL University, Sorbonne Universit\'e and Universit\'e de
  Paris, 75005 Paris, France}
\affiliation{\lastequal}

\begin{abstract}
In many applications in biology, engineering and economics, identifying similarities and differences between distributions of data from complex processes requires comparing finite categorical samples of discrete counts.  Statistical divergences quantify the difference between two distributions.  However their estimation is very difficult and empirical methods often fail, especially when the samples are small.  We develop a Bayesian estimator of the Kullback-Leibler divergence between two probability distributions that makes use of a mixture of Dirichlet priors on the distributions being compared.  We study the properties of the estimator on two examples: probabilities drawn from Dirichlet distributions, and random strings of letters drawn from Markov chains.  We extend the approach to the squared Hellinger divergence.  Both estimators outperform other estimation techniques, with better results for data with a large number of categories and for higher values of divergences.

\end{abstract}

\maketitle

\section{Introduction}
{\IN Understanding of the structure and function of a large number of biological systems requires comparison between two probability distributions of their states or activities, generated under different conditions.  For example, one may be interested in how the distribution of neural firing patterns underlying typical vocalizations in a song bird is different from patterns used to drive atypical, exploratory vocal behaviors~\cite{hernandez_unsupervised_2022}.  One can similarly ask how different are the distributions of stimuli encoded by two different firing patterns; the difference then can be viewed as a measure of semantic similarity between these patterns \cite{ganmor_thesaurus_2015}.  In the context of immunology, one is often interested in information theoretic quantities in order to quantify diversity or to assess differences between distributions of immune receptors \cite{mora_quantifying_2019,camaglia_quantifying_2023}.  In these and similar examples, the Kullback-Leibler (KL) divergence {\DKL}, also known as relative entropy, is often used as a measure of dissimilarity.  It is a non-symmetric measure of the difference between two probability distributions with a wide range of applications in information theory~\cite{kullback_information_1951}.  While not a distance in the mathematical sense, it is often the choice measure of dissimilarity since it can be applied to categorical (non-ordinal) data, when the usual statistical moments such as the mean and variance are not well defined.  Indeed, like other ``information theoretic quantities'', the KL divergence is not associated to the category itself, but rather to the underlying probability distribution~\cite{zhang_entropic_2022}.}

{\IN Estimation of information theoretic quantities is a hard problem, with a lot of attempts in the recent literature.  Most of these have focused on the entropy and mutual information, but estimation of the KL divergence has also been investigated~\cite{zhang_nonparametric_2014}.  When faced with data without any knowledge of the true underlying distribution, empirical approaches (typically referred to as ``naive'' \cite{strong_entropy_1998} or ``plugin'' \cite{antos_convergence_2001}) are often used. } These methods approximate the true probabilities of events with their empirical frequencies, with an optional pseudocount.  These types of estimators have been investigated thoroughly.  The consensus is that, for all entropic quantities, these estimates are typically strongly biased~\cite{antos_convergence_2001,paninski_estimation_2003,zhang_entropy_2012,jiao_maximum_2015}.  To overcome this limitation, other approaches have been proposed to estimate the Shannon entropy (or the mutual information) of categorical data.  These techniques include Bayesian methods~\cite{nemenman_entropy_2001, archer_bayesian_2014}, coverage adjusted methods~\cite{chao_nonparametric_2003} and bias corrected methods~\cite{paninski_estimation_2003, schurmann_bias_2004, zhang_entropy_2012}.  In the case of the KL divergence, the cross-entropy term, which diverges due to contributions where one distribution has samples and the other does not, makes it difficult to extend these methods in the absence of information about the joint distribution.  The bias-corrected ``Z-estimator''~\cite{zhang_nonparametric_2014}, proposed for KL divergence estimation, tackles these issues. However, it has a strong dependence on the sample size.

Here we propose a Bayesian estimator of the DKL for systems with finite number of categories using a mixture of symmetric Dirichlet priors (Dirichlet Prior Mixture, or DPM).  {\IN This approach is the generalization of the main idea from~\cite{nemenman_entropy_2001} that, to produce unbiased estimators, one needs to start with Bayesian priors that are (nearly) uniform not on the space of probability distributions, but directly on the quantity being estimated.  Here we extend this idea beyond the estimation of entropy, for which it was first developed.} We check that, for data distributed according to a Dirichlet prior, our new approach for estimation of the KL divergence consistently converges faster to the true value than other methods.  We provide an algebraically equivalent expression for the Z-estimator (following~\cite{schurmann_note_2015}), which makes it applicable to large sample sizes.  We also test the DPM technique on sequences generated by Markov chains, which are not typical within the DPM prior, obtaining better performance for datasets with many categories.  {\IN We then focus our analysis on another measure of similarity between categorical distributions, the Hellinger divergence~\cite{hellinger_neue_1909}, which, unlike the DKL, is a well defined bounded distance between distributions.  To show the generality of our approach, we also develop a DPM estimator for the squared Hellinger divergence.  In computational tests, we show the DPM approach to be accurate for this quantity as well.  Since no estimation method can be guaranteed to estimate entropic quantities without a bias for an arbitrary underlying probability distribution, we finish by discussing the method's reliability when applied to real experimental data, where the true values of the divergences are not known a priori.  }

\section{Results}
\subsection{Bayesian framework for the estimation of the divergence}

\begin{figure}
\begin{center}
\includegraphics[width=\columnwidth]{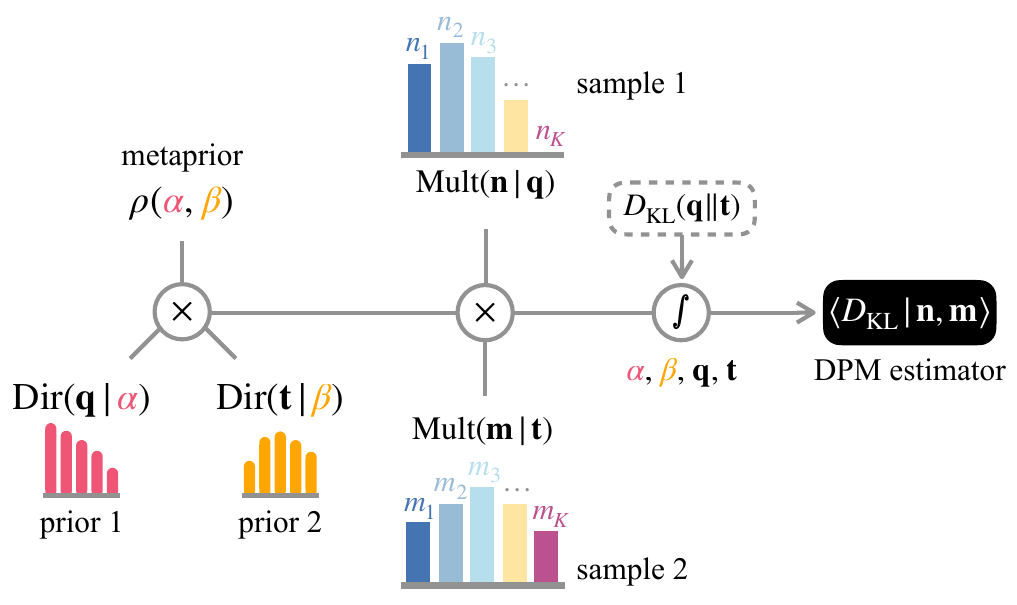}
\caption{
Schematic representation of the Bayesian approach for the inference of the Kullback-Leibler divergence.
Given two independent samples $\vec{n}$ and $\vec{m}$ of categorical
data, we model the true distribution $\vec q$ and $\vec t$ as drawn
from a mixture of Dirichlet distribution with unknown concentration
parameters $\alpha$ and $\beta$. The DPM estimation of the {\DKL} is then obtained by
averaging over all values of these parameters, weighted by the
likelihood of the samples $\vec n$ and $\vec m$ (multinomial
distributions).
}
\label{fig:initial_scheme}  
\end{center}
\end{figure}

Our goal is to derive an estimate of the Kullback-Leibler divergence between the distributions of categorical variables $\vec{t}$ and $\vec{q}$, $\DKL(\vec{q}\Vert\vec{t})$.
We consider a discrete set of $K$ categories labeled with $i=1,\dots,K$. 
Examples of categorical variables include ``words" defined as
sequences of neuron firing patterns (spike counts in time windows),
sets of coexisting ecological or molecular species or a sequence of
amino acids or nucleotides.  Each category $i$ has a certain (unknown)
probability $q_i$ in the first condition, and $t_i$ in the second
condition.  We observe this category $n_i$ times in an experiment done
in the first condition, and collect the data in the histogram
$\vec{n}=\lbrace n_i\rbrace_{i=1}^K$.  An experiment in the second
condition returns the counts $\vec{m}=\lbrace m_i\rbrace_{i=1}^K$.  We
want to estimate the Kullback-Leibler divergence between $\vec{t}$ and
$\vec{q}$~\cite{kullback_information_1951}, defined as:
\begin{equation}
\DKL(\vec{q}\Vert\vec{t}) = H(\vec{q}\Vert\vec{t})-S(\vec{q}) =\sum_{i=1}^K q_i \log \frac{q_i}{t_i},
\label{eq:DKL0}
\end{equation}
where {\IN we defined}  the cross
entropy between $\vec{t}$, and $\vec{q}$, 
$H(\vec{q}\Vert\vec{t}) = - \sum_{i=1}^K q_i \log t_i$,  and the Shannon
entropy,
$S(\vec{q}) = - \sum_{i=1}^K q_i \log q_i$~\cite{shannon_mathematical_1948}.

Taking inspiration from Nemenman et al.~\cite{nemenman_entropy_2001},
we choose to estimate the $\DKL$ in a Bayesian framework.
The approach is summarized in Fig.~\ref{fig:initial_scheme}.  We do
not have access to the true probability distributions $\vec{t}$ and
$\vec{q}$, only to the empirical histograms $\vec{n}$ and $\vec{m}$.
The simple method consisting in approximating
$q_i\approx n_i/\sum_jn_j$ and likewise for $\vec t$ into
Eq.~\ref{eq:DKL0} is known to work very
poorly~\cite{zhang_entropy_2012,jiao_maximum_2015}.  {\IN The issue
  comes from the presence of categories never observed in one sample,
  while they are present in the other, resulting in divergence of the
  logarithmic term.  } To go beyond that, we construct a prior of the
true distributions $P_{\rm prior} (\vec{q}, \vec{t})$ and weight the
estimate of the divergence by posterior over $\vec q$ and $\vec t$:
\begin{equation}
\langle \DKL (\vec{q}, \vec{t}) \vert \vec{n},\vec{m} \rangle
=
\int \! {d\vec{q}} {d\vec{t}} \; P_{\rm post} (\vec{q}, \vec{t}) \DKL(\vec{q}\Vert\vec{t})
,
\label{DKL1}
\end{equation}
where
\begin{equation}
P_{\rm post} (\vec{q}, \vec{t})
=
\frac{1}{Z}P_{\rm prior} (\vec{q}, \vec{t}) P(\vec{n}, \vec{m} \vert \vec{q}, \vec{t})
,
\label{post}
\end{equation}
with $Z=P(\vec n,\vec m)= \int \! {d\vec{q}} {d\vec{t}} \; P_{\rm prior} (\vec{q},\vec{t}) P(\vec{n}, \vec{m}|\vec{q}, \vec{t})$ a normalization.

The empirical observations $\vec n$ and $\vec t$ are assumed to be independent samples of $\vec q$ and $\vec t$ respectively, and are thus distributed according to a multinomial distribution:
\begin{equation}
P( \vec{n}, \vec{m} \vert \vec{q},\vec{t} )
=
\Mult{n}{q} \Mult{m}{t},
\end{equation}
with
\begin{equation}
\label{eqn:multinomial}
\Mult{n}{q}=
\frac{N !}{\prod_{i=1}^K n_{i} !}
\prod_{i=1}^K {q_i} ^{n_i}
=
\frac{1}{B(\vec{n}+\vec{1})}
\prod_{i=1}^K {q_i} ^{n_i},
\end{equation}
where $N=\sum_{i=1}^K n_{i}$, and $B(\vec x)$ is the multivariate Beta function: $B(\vec{x})=\prod_{i=1}^K \Gamma (x_{i})/{\Gamma(\sum_{i=1}^K x_{i})}$, where $\Gamma(x)$ is the gamma function. 

A natural choice for the prior on $\vec q$ and $\vec t$ is {\IN the} Dirichlet distribution, which is the conjugate prior of the multinomial distribution, and is defined as
\begin{equation}
\label{eqn:Dirichlet-prior}
\Dir{q}{\alpha} = \frac{\delta\left(\sum_{i=1}^Kq_i-1\right)}{B(\vec{\alpha})} \prod_{i=1}^K {q_{i}} ^{\alpha-1},
\end{equation}
{\IN where $\alpha \in (0,\infty)$ is the ``concentration parameter'',
  $\vec{\alpha}=\lbrace\alpha\rbrace_{i=1}^{K}$ and $\delta(x)$ is
  {\IN the} Dirac's delta function imposing normalization.  Rank plots
  associated to $\Dir{q}{\alpha}$ are shown in
  Fig.~\ref{fig:dirichlet-prior}A.  }
For $\alpha\to \infty$, the prior tends to a uniform distribution $q_i=1/K$.
For small concentration parameters $\alpha$, the distribution is peaked with weights given to just a few categories.

As noted in Ref.~\cite{nemenman_entropy_2001}, entropies of
distributions drawn from a Dirichlet with the same $\alpha$ all have
similar entropies, strongly biasing the Shannon entropy estimate,
especially in the large $K$ limit.  {\IN To reduce the bias, one then
  uses a mixture of Dirichlet distributions at different $\alpha$,
  allowing substantially different values of the entropy \textit{a
    priori}.  For a certain choice of the mixture distribution (the
  prior over $\alpha$, $\rho(\alpha)$), one can achieve a
  nearly-uniform \textit{a priori} distribution of entropies and,
  consequently, a much smaller estimation bias
  \cite{nemenman_entropy_2001,hernandez_low-probability_2023}.  }
We expect {\DKL} also to have very similar values for all
distributions generated from the Dirichlet priors with fixed $\alpha$
and $\beta$.  We then expect that a good estimator may be produced by
using a mixture of Dirichlet
distribution that allows to span different values of the expected
{\DKL}:
\begin{equation}
\label{eqn:new-star-prior}
P_{\rm prior}( \vec{q},\vec{t})=
\int_0^\infty \! \int_0^\infty \! {d\alpha} {d\beta} \;
\rho(\alpha,\beta) \Dir{q}{\alpha} \Dir{t}{\beta}
,
\end{equation}
where $\rho(\alpha,\beta)$ is a {\IN ``hyper-prior'', i.~e., a prior over the hyper parameters $\alpha$ and $\beta$}.
Plugging this prior into Eq.~\ref{DKL1} and~\ref{post} gives:
\begin{align}
\label{eqn:DKL-estimator}
\nonumber
&\langle \DKL \vert \vec{n},\vec{m} \rangle 
=\\
&=
\frac{1}{Z} 
\int \! {d\alpha} {d\beta} \;
P\left( \vec{n}|\alpha\right)P\left(\vec{m}\vert\beta\right)\rho(\alpha,\beta)
\langle \DKL \vert \vec{n},\vec{m} ; \alpha,\beta\rangle
,
\end{align}
with
\begin{align}
\label{likelihood}
&P\left( \vec{n}\vert\alpha\right)
= \int\! {d\vec{q}} \,\Mult{n}{q} \,\Dir{q}{\alpha}
=
\frac{B(\vec n+\vec \alpha)}{B(\vec\alpha)B(\vec n+\vec 1)}
\end{align}
and likewise for $P(\vec m|\beta)$.
The normalization now reads
\begin{equation}
Z = \int \! {d\alpha} {d\beta} \; 
P( \vec n, \vec m \vert \alpha, \beta)
.
\end{equation}
The expected value of the {\DKL} inside the integral may be computed analytically (see {\apdx}~\ref{AppDirMult}): 
\begin{align}
\label{eqn: posterior-DKL}
\nonumber
&\langle \DKL \vert \vec{n},\vec{m} ; \alpha,\beta\rangle 
=\\
\nonumber
&=
\int\! {d\vec{q}} {d\vec{t}} P(\vec q,\vec t|\vec m,\vec
n,\alpha,\beta)\DKL(\vec q\Vert\vec t)
\\
\nonumber
&=
\sum_i\frac{n_i+\alpha}{N+K\alpha} 
\lbrace \Delta\psi(M+K\beta , m_i+\beta) 
\\
& 
-\Delta\psi(N+K\alpha+1 , n_i+\alpha+1) \rbrace
.
\end{align}
where $\Delta\psi(z_1,z_2)=\psi(z_1)-\psi(z_2)$ is the difference of digamma functions $\psi$ ({\IN i.~e.}, polygamma function of order $0$, see Eq.~\ref{eqn:polygamma}).

\begin{figure}
\begin{center}
\includegraphics[width=\columnwidth]{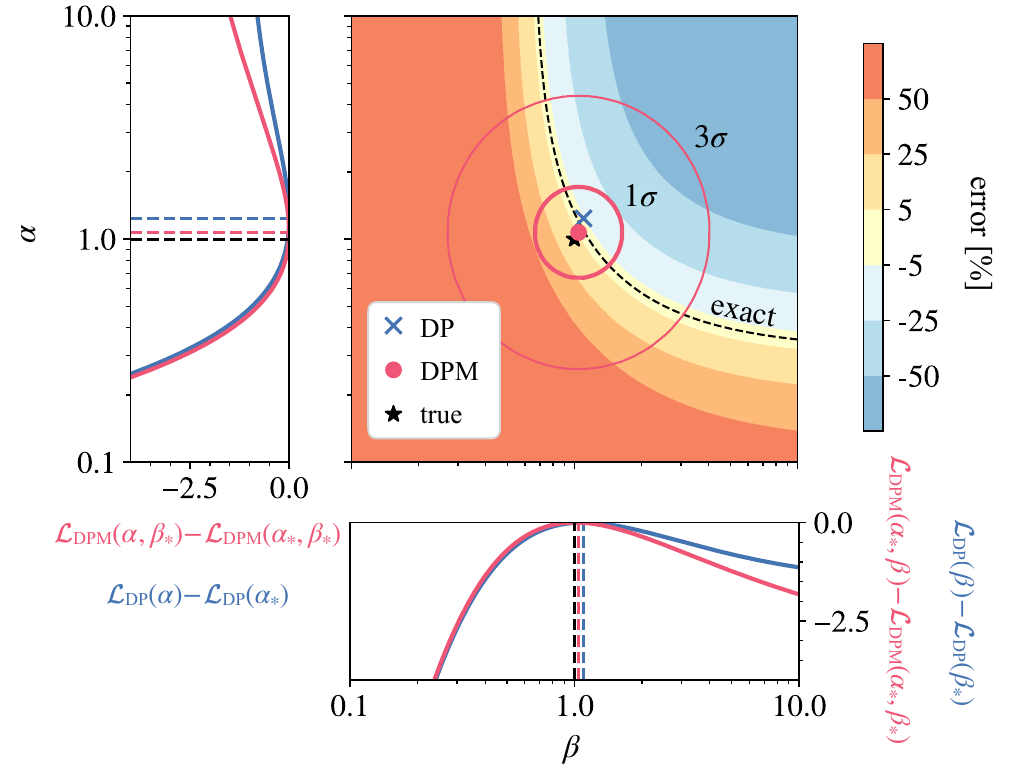}
\caption{
Dependency on the concentration parameters $\alpha$ and $\beta$ of the two main factors appearing in the average performed by
the DPM estimator (Eq.~\ref{eqn:DKL-estimator}): the posterior
$\propto P\left(
  \vec{n}|\alpha\right)P\left(\vec{m}\vert\beta\right)\rho(\alpha,\beta)$,
and the expected value $\langle \DKL \vert \vec n, \vec m ; \alpha,
\beta \rangle$.
For reasons of accuracy, integrals are numerically computed
in logarithmic space, so that it's more informative to introduce the
posterior density in $\log\alpha$ and $\log\beta$, so that we define:
$
\mathcal{L}_{\rm DPM}(\alpha,\beta)
\equiv
\log_{10} P(\bm{n} \vert \alpha) P(\bm{m} \vert \beta)\rho(\log\alpha,\log\beta)
$.
We compare it with the log-likelihoods
$\mathcal{L}_{\rm DP}(\alpha)\equiv P(\vec n \vert \alpha)$ and
$\mathcal{L}_{\rm DP}(\beta) \equiv P(\vec m|\beta)$ in the left and bottom subplots.
The central panel shows the relative error associated to $\langle \DKL
| \vec{n},\vec{m}; \alpha,\beta \rangle$ as a function of the two
concentration parameters $\alpha$ and $\beta$. The samples $\vec{n}$ and
$\vec{m}$ of sizes $N=M=\frac{1}{4}K$ were generated from two distinct Dirichlet-multinomial processes with concentration parameters $\alpha_{\rm true}=1.0$ and $\beta_{\rm true}=1.0$ with $K=400$ (black star on the central panel). 
The dashed black line corresponds to 0 error.
Blue cross: maximum of $\mathcal{L}_{\rm DP}(\alpha)+\mathcal{L}_{\rm DP}(\beta)$.
Red circle: maximum of $\mathcal{L}_{\rm DPM}(\alpha,\beta)$.
Red lines are standard deviations associated to $\mathcal{L}_{\rm DPM}$ around its maximum. 
}
\label{fig:evidence}
\end{center}
\end{figure}

Similarly we can calculate
$\langle \DKL^2 \vert \vec{n},\vec{m}\rangle$, which we can use to
compute the {\IN posterior} standard deviation of our method
({\apdx}~\ref{AppDirMult}).  {\IN For a given choice of
  $\rho(\alpha,\beta)$, the DPM estimate for $\DKL$ in
  Eq.~\ref{eqn:DKL-estimator} can be computed numerically (same for
  $\DKL^2$ in Eq.~\ref{eqn:squared-DKL-estimator}), as described in
  detail in {\apdx}~\ref{AppNumImp}. }
The code is available on github as specified in {\apdx}~\ref{AppCodAva}.

We expect that, in the limit of large data ($N,M\gg K$), the integral
of Eq.~\ref{eqn:DKL-estimator} will be dominated by the values of
$\alpha$ and $\beta$ that maximize the likelihoods
$P(\vec{n}\vert\alpha)$ and $P(\vec{m}\vert\beta)$, regardless of
the 
{\IN hyper-prior} $\rho(\alpha,\beta)$.
{\IN The dominant role of the likelihood $P(\vec{n}\vert\alpha)$ for
  increasing $N$ was equivalently observed for the NSB entropy
  estimator~{\cite{nemenman_coincidences_2011}}.  } By contrast, we
expect the {\IN prior} $\rho(\alpha,\beta)$ to play a role in the
low-sampling regime, as can be seen from Fig.~\ref{fig:evidence}.

{\IN A simplified approach for the estimation of the $\DKL$ would then
  be to provide a choice for the concentration parameters that
  maximizes the likelihoods $P(\vec n \vert\alpha)$ and
  $P(\vec m \vert\beta)$ (see Eq.~\ref{likelihood}).}  We refer to the
application of Eq.~\ref{eqn: posterior-DKL} with the
maximum-likelihood values of $\alpha$ and $\beta$
as the Dirichlet Prior (DP) estimator.

\subsection{Choosing the {\IN hyper-prior}}

\begin{figure}
\begin{center}
\includegraphics[width=\columnwidth]{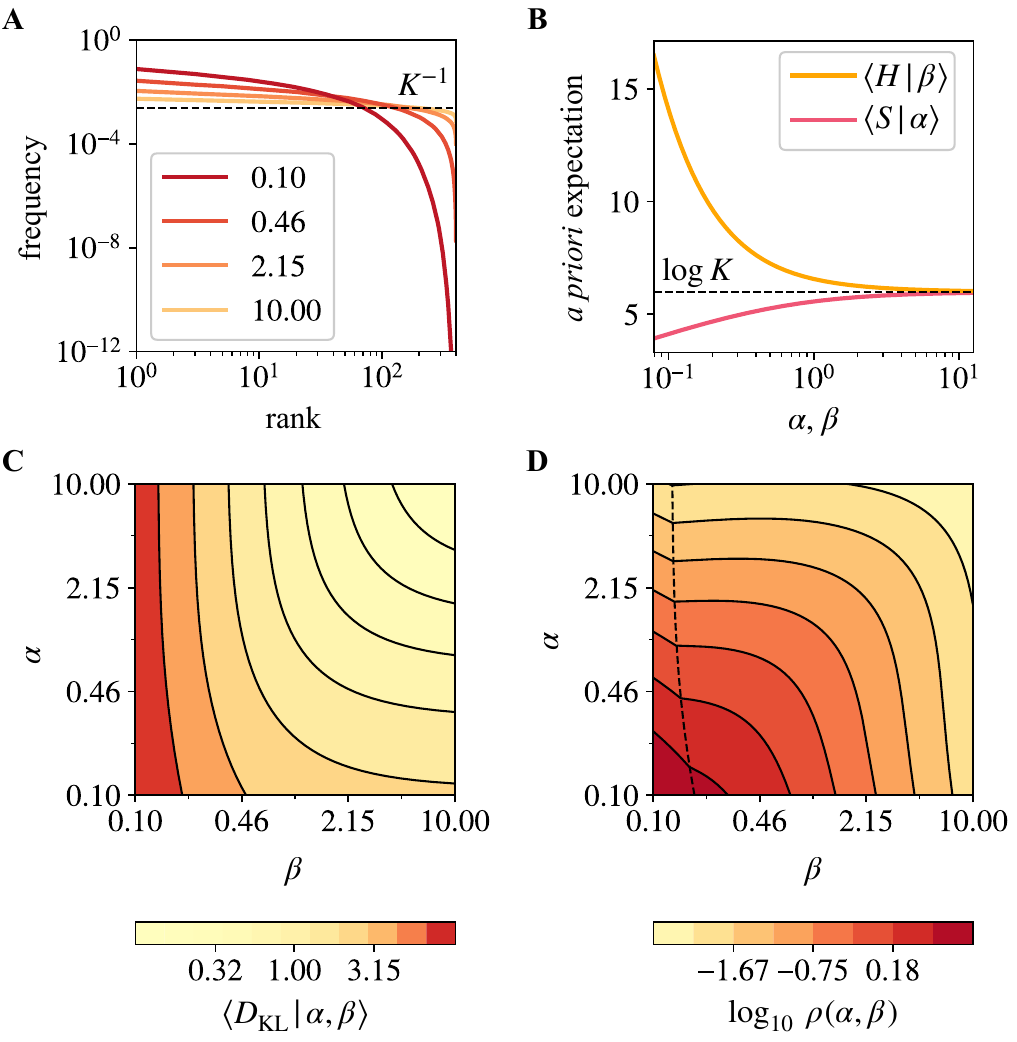}
\caption{
\textbf{A.} Average rank-frequency plots for probabilities drawn from
a Dirichlet prior $\Dir{q}{\alpha}$ for different choices of
concentration parameters $\alpha$.
\textbf{B.} Expected values of the cross-entropy $\langle H | \beta \rangle$ and the entropy $\langle S | \alpha \rangle$ under Dirichlet priors as functions of the concentration parameters. 
\textbf{C.} Expected value of the {\DKL} divergence under Dirichlet priors $\langle \DKL | \alpha, \beta \rangle$ as a function of the two concentration parameters. 
\textbf{D.} Log-metaprior $\log_{10} \rho(\alpha,\beta)$ as a function of the two concentration parameters. Dashed black line represents the level $\langle \DKL | \alpha, \beta \rangle = \log K$.  $K=20^2$ for all plots. 
}
\label{fig:dirichlet-prior}  
\end{center}
\end{figure}

To finalize the $\DKL$ estimation, we need to choose a functional form for the hyper-prior $\rho(\alpha,\beta)$ in such a way that the resulting ensemble has an evenly distributed {\DKL}. 
In the limit of large numbers of categories ($K\gg 1$), both contributions of the {\DKL}, $S(\vec q)$ and $H(\vec q\Vert t)$ are very peaked around their mean values, which can be computed analytically ({\apdx}~\ref{AppDirMult}):
\begin{equation}
\mathcal{A}(\alpha)\equiv\langle S \vert \alpha\rangle=\Delta
\psi(K\alpha+1,\alpha+1)\leq \log K,
\end{equation}
and
\begin{equation}
\mathcal{B}(\beta) \equiv \langle H \vert \alpha,\beta \rangle=\Delta \psi(K\beta,\beta) \geq \log K
\end{equation}
(which only depends on $\beta$), at fixed concentration parameters.
These mean values are shown in Fig.~\ref{fig:dirichlet-prior}B, and
the corresponding $\DKL=H-S$ in Fig.~\ref{fig:dirichlet-prior}C as a
function of $\alpha$ and $\beta$.
{\IN We are interested in finding a hyper-prior such that the
  resulting prior over $\DKL$ is not peaked.  This results in the
  following inverse problem for finding the hyper-prior $\rho_{z}(z)$,
  where we denote $\DKL$ by $z$:
\begin{equation}
\label{eqn:marginalization}
\rho_{z}(z) \approx 
\int_0^\infty \! {d\alpha} \int_0^\infty \! {d\beta} \;
\rho(\alpha,\beta) \, \delta\left( \mathcal{B}(\beta) - \mathcal{A}(\alpha)- z \right),
\end{equation}
with the choice $\rho_{z}(z)$ to be made.}  Because we have a
one-dimensional target distribution $\rho_z(z)$, but a 2-dimensional
hyper-prior $\rho(\alpha,\beta)$, there are infinitely many solutions
to this inverse problem.  Without losing generality, we can make the
change of variable from $\alpha$ and $\beta$ to $\mathcal{A}$ and
$\mathcal{B}$:
\begin{equation}
\label{eqn:marginalization-pt2}
\rho_z(z) \approx 
\int_0^{\log K} \! {d\mathcal{A}}\int_{\log K}^{+\infty} {d\mathcal{B}} \;
\rho_{AB}(\mathcal{A},\mathcal{B}) \, \delta\left( \mathcal{B} - \mathcal{A}- z \right),
\end{equation}
with
\begin{equation}
\label{phi1}
\rho(\alpha,\beta)=\left\vert\partial_{\alpha} \mathcal{A}\right\vert
\left\vert\partial_{\beta}
  \mathcal{B}\right\vert\rho_{AB}(\mathcal{A}(\alpha),\mathcal{B}(\beta)).
\end{equation}
Then a natural choice is to pick the Ansatz imposing that all values of $\mathcal{A}$ and $\mathcal{B}$ with the same {\DKL} are equiprobable:
\begin{equation}
\label{phi2}
\rho_{AB}(\mathcal{A},\mathcal{B})=\phi(\mathcal{B}-\mathcal{A}).
\end{equation}
Then $\phi(z)$ satisfies:
\begin{align}
\label{phi0}
\nonumber
\rho (z) &=
\phi(z) \int_0^{\log K} \! {d\mathcal{A}} \; \theta( z + \mathcal{A} -\log K )
\\
&=
\phi(z) \lbrace z\; \theta(\log K - z) + \log K \; \theta(z - \log K) \rbrace
,
\end{align}
where $\theta(x)=1$ if $x\geq 1$ and $0$ otherwise (Heaviside function), or after inversion:
\begin{equation}
\label{eqn:meta_prior-phi_func}
\phi(z) =
\begin{cases}
\rho(z) {z}^{-1} \qquad &z < \log K \\
\rho(z) \dfrac{1}{\log K} \qquad &\text{otherwise}.
\end{cases}
\end{equation}
Eqs.~\ref{phi1}, \ref{phi2}, and~\ref{eqn:meta_prior-phi_func} give us
the final form of the {\IN hyper-prior} $\rho(\alpha,\beta)$.  We are
left with the choice of the distribution of the {\DKL}, $\rho(z)$.
{\IN We pick a log-uniform (also known as ``reciprocal'')
  distribution,
  $\rho(z)\propto z^{-1}$~\cite{hamming_distribution_1970}, allowing
  to evenly span over different orders of magnitude of the $\DKL$.}
The resulting hyper-prior is represented in
Fig.~\ref{fig:dirichlet-prior}D.

\subsection{Tests on synthetic Dirichlet samples }

\begin{figure}
\begin{center}
\includegraphics[width=\columnwidth]{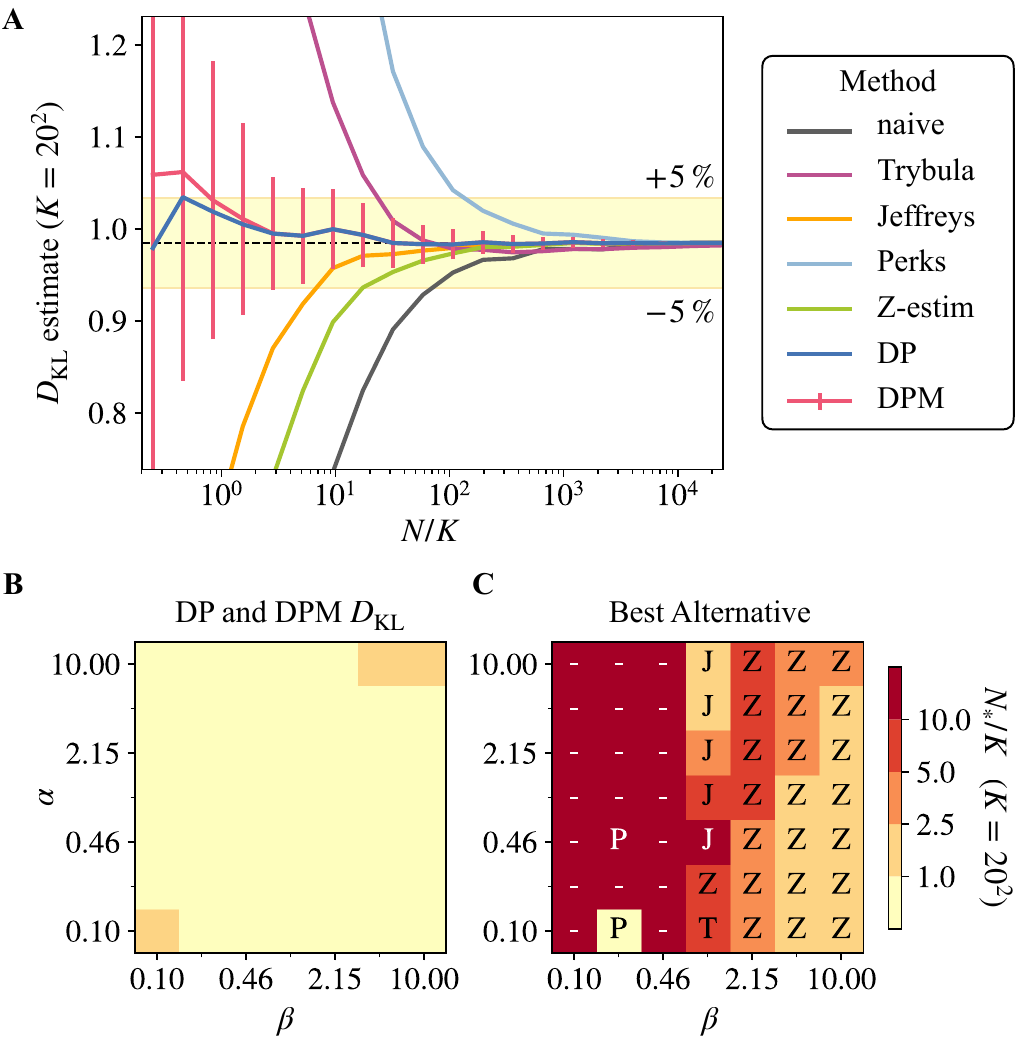}
\caption{
Convergence of the $\DKL$ estimates for increasing sample sizes.
\textbf{A.}
We draw two independent histograms from Dirichlet-multinomial
distributions with parameters $\alpha$ and $\beta$.
We obtain subsamples of different sizes $N=M$ and we estimate the
{\DKL} divergence for each of them.
We compare the DP and DPM results to those obtained with the known
alternative estimators (Table~\ref{tab:plugin-estimators} and
Eq.~\ref{eqn:z-estimator-final}), as a function of $N/K$.
Here we plot the average over 100 repetitions for concentration parameters $\alpha=\beta=1$ and $K=20^2$.
The highlighted region in yellow corresponds to an error of $\pm 5\%$ relative to the average true value, represented by the dashed black line.
\textbf{B.}
Convergence of {\DKL} estimators for different (log-spaced) concentration parameters $\alpha$, $\beta$. 
We plot the $N_{\ast}/K$ score for the size at which the best between the DP and DPM estimators reach the true value up to a relative error of $\pm 5\%$ (highlighted region in A).
Lower $N_{*}/K$ scores correspond to faster converge of the estimator. 
\textbf{C.}
The method with the best convergence score among the alternative methods is represented (first letter of its name).
A dash symbol ``-'' indicates that no alternative has a score  $N_{\ast} / K < 50$.
The DP and DPM estimators shows faster convergence compared to all
other methods for all parameters. 
}
\label{fig:dirichlet-convergence-estimation}
\end{center}
\end{figure}

To assess the properties of the DPM estimator, we test it on data
generated from distributions drawn from Dirichlets
$\vec{q} \sim \Dir{q}{\alpha}$, $\vec{t} \sim \Dir{q}{\beta}$
(Eq.~\ref{eqn:Dirichlet-prior}), for various values of $\alpha$ and
$\beta$.  Having in mind applications to polypeptide sequences, we
perform our tests for three different numbers of categories $K=20^2$,
$20^3$, and $20^4$, {\IN the numbers of all possible 2-mers, 3-mers
  and 4-mers that can be produced with an alphabet of 20 letters
  (e.~g., amino acids).} For each choice of $\vec q$ and $\vec t$,
samples $\vec n$ and $\vec m$ are generated from these distributions.
{\IN This application may be viewed as a the consistency check for te
  estimator, since the estimator relied on the Dirichlet hypothesis,
  which is satisfied by the data.  }

{\IN We know that standard Bayesian consistency applies, ensuring that
  DPM (and DP) estimators converge to the true value in the limit of
  large samples.  To understand how DPM estimator converges to the
  true value, we extract subsamples of increasing sizes $N=M$ from a
  larger sample of size $2\cdot 10^7$.}
Fig.~\ref{fig:dirichlet-convergence-estimation} compares our {\DKL}
estimate to several state-of-the-art estimators: the additive
smoothing method with different values of the pseudo-count (see below
for details), the Z estimator, and the simplified version of our
method, the DP estimator, obtained by fixing $\alpha$ and $\beta$ to
their maximum-likelihood values.

Additive smoothing estimators are defined as: $\DKL(\vec{\hat q}\Vert\vec{\hat t})$, with $\hat q_i=(n_i+a)/(N+Ka)$, and $\hat
t_i=(m_i+b)/(M+Kb)$.
We use 4 choices for the pseudo-counts $a$ and $b$, summarized in Table~\ref{tab:plugin-estimators}.
To avoid infinite values, in the case $b=0$ we set to zero the terms for which $m_i=0$.

\begin{table}[ht]
\begin{tabular}{ |c|c|c|c| } 
\hline
name & $a$ & $b$ & reference \\
\hline
``naive'' & $0$ & $0$ & - \\
``Jeffreys'' & $0.5$ & $0.5$ & \cite{jeffreys_invariant_1946} \\
``Trybula'' & $\sqrt{N} / K$ & $\sqrt{M} / K$ & \cite{trybula_problems_1958} \\
``Perks'' & $1 / K^{(\rm obs)}$ & $1 / K^{(\rm obs)}$ & \cite{perks_observations_1947} \\
\hline
\end{tabular}
\caption{List of choices for the pseudo-counts used to define alternative estimators of the {\DKL}~\cite{hausser_entropy_2009}.
$K^{(\rm obs)} \leq K$ is the number of observed categories, {\IN for which} $n_i>0$ in each distinct sample.
}
\label{tab:plugin-estimators} 
\end{table}

{\IN It has been shown that naive estimators converge to the true
  value in the limit of large samples, but have an infinite bias due
  to low-probability categories~\cite{zhang_nonparametric_2014}.}  The
``Z-estimator''~\cite{zhang_nonparametric_2014} was introduced to
remove this bias asymptotically.  Although its original definition was
given as a series, one can show following~\cite{schurmann_note_2015}
that its expression reduces to (Appendix~\ref{AppZhang}):
\begin{equation}
\label{eqn:z-estimator-final}
\widehat{D}_{\rm KL}^{(Z)}
=
\sum_{i=1}^K \frac{n_i}{N}
\left[
\Delta \psi (M+1, m_{i}+1)-\Delta \psi (N, n_{i})
\right],
\end{equation}
where the first term in the sum corresponds to an estimator of
$H(\vec q\Vert t)$, and the second term is the classic
Schurmann-Grassberger estimator of the entropy
$S(\vec q)$~\cite{schurmann_bias_2004}.  In Appendix~\ref{AppZhang} we
observe that
$\langle \DKL \vert \vec n, \vec m, \alpha, \beta \rangle \to
\widehat{D}_{\rm KL}^{(Z)}$ in the limit $\alpha \to 0$,
$\beta \to 1$, $N \gg K$ and $M \gg K$.

Comparing the convergence of the different estimators to the true
{\DKL} value as a function of the subsample size $N/K$ for
$\alpha=\beta=1$ and $K=20^2$
(Fig~\ref{fig:dirichlet-convergence-estimation}A), we see that the DPM
performs better than other estimators.  To assess how performance
depends on the concentration parameters, we repeated this convergence
analysis for different values of $\alpha$ and $\beta$.  We measure
convergence through $N^{\ast}$, defined as the sample size where the
estimator get within $5\%$ of the true value
(Fig~\ref{fig:dirichlet-convergence-estimation}B).  {\IN This measure
  of accuracy has the advantage to be applicable to all considered
  estimation methods.  }

{\IN Our estimator consistently performs well and compares favorably
  to other methods when data was generated from distributions drawn
  from symmetric Dirichlet.  In most cases, the proposed DPM estimator
  converges faster than all other considered methods
  (Fig~\ref{fig:dirichlet-convergence-estimation}C).  }
{\IN The better performance is striking also for larger numbers of
  categories, $K=20^3$ and $20^4$
  (Fig.~\ref{fig-si:dkl-dirichlet-larger-K}).  }

\subsection{Tests on synthetic Markov chain sequences }

\begin{figure}
\begin{center}
\includegraphics[width=\columnwidth]{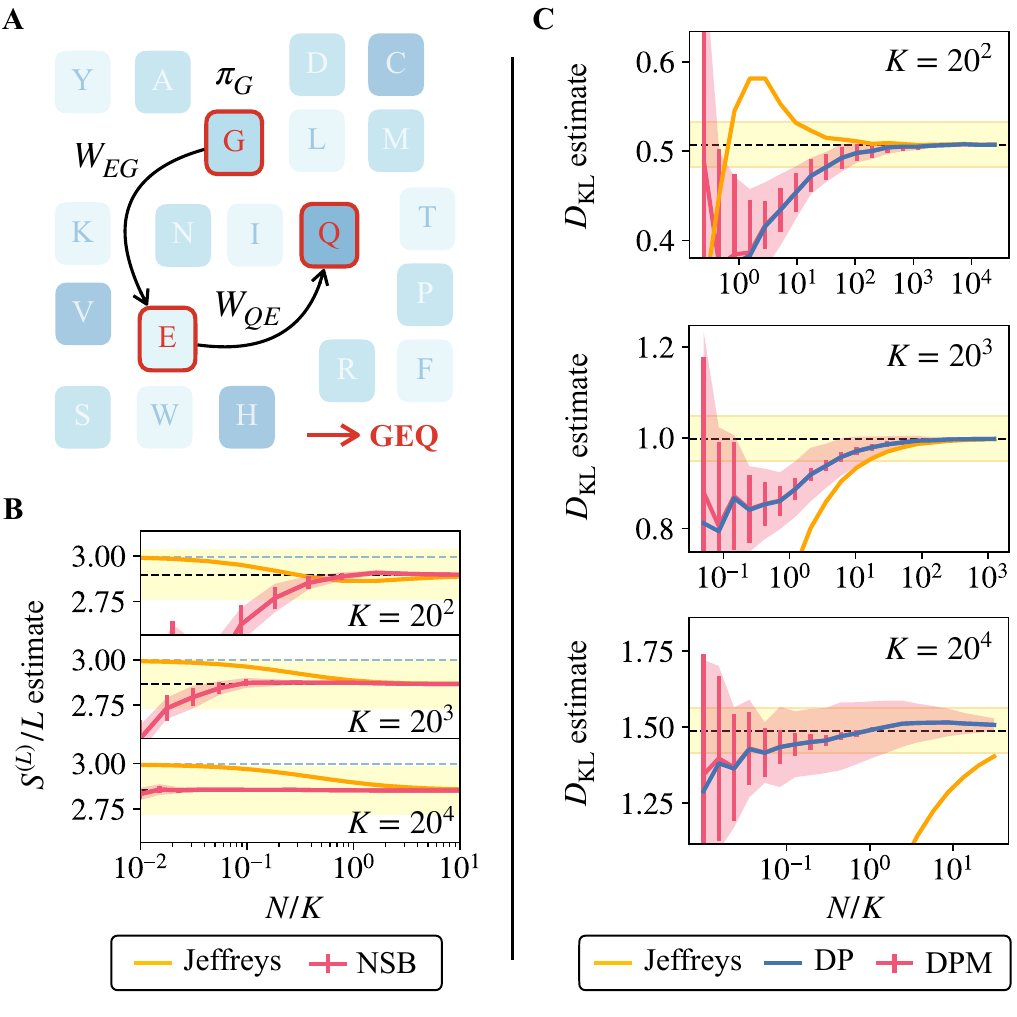}
\caption{
\textbf{A.} Schematic representation of the generation process of a $L=3$-gram using a Markov chain with $20$ states. 
\textbf{B.} 
We generate a random Markov matrix and draw a sample of $2\cdot10^8$ independent $L$-grams, for $L=2,3,4$.
The Shannon entropy $S^{(L)}$ (Eq.~\ref{eqn:entropy_Lgram_Markov}) estimated with the NSB method \cite{nemenman_entropy_2001} for sub-samples of size $N$, averaged over 30 repetitions normalized by the true value of the entropy and rescaled by the asymptotic value.
The highlighted yellow region corresponds to a $\pm 5 \%$ error range with respect to the average true value (dashed black line).
\textbf{C.}
{\DKL} estimate and its standard deviation as a function of relative subsample size $N/K$, for $K=20^2, 20^3, 20^4$.
For $K=20^2$, the DP and DPM estimators perform comparably to the best alternative (``Jeffreys'' for all $K$), while {\FC they work} better for larger $K$.
The error bars represent the average {\FC posterior} standard deviation of the {\DKL} estimate associated to the DPM method.
The red shade is the standard deviation of the DPM {\DKL} estimates across the repetitions.  
}
\label{fig:markov-Lgrams}
\end{center}
\end{figure}

To test the performance of DPM on a different synthetic system that
does not satisfy the Dirichlet assumption, we generated $L$-long
sequences (or ``$L$-grams'') from a Markov chain described by the
transition matrix $\hat{W} \in \mathcal{M}_{20}$ with 20 states
$\mu=1,\cdots,20$.  We choose each transition probability
$P(\mu\to\nu)$ from a uniform distribution in $(0,1)$ and then impose
that the transition matrix is a right stochastic matrix,
$P(\mu\to\nu)=W_{\nu\mu}$ {\IN by normalizing to 1 each column of the
  transition matrix.  An illustration where the states are the 20
  amino acids is shown in Fig.~\ref{fig:markov-Lgrams}A.}
With this choice for the Markov transition matrix, all states
communicate and are non absorbing.
We verify there exists a stationary probability vector {\IN
  $\vec{\pi} = \lbrace \pi_{\mu} \rbrace_{\mu=1}^{L}$} that satisfies
$\vec{\pi} = \hat{W} \vec{\pi}$.  The number of categories is $K=20^L$
and each category $i$ corresponds to the $L$-gram
$(x_{1},\cdots,x_{L})$ with the stationary probability $q_i$ equal to
$q_i=\pi_{x_{1}} W_{x_{2} x_{1}} \cdots W_{x_{L} x_{L-1}}$.
 
We analytically compute the entropy associated to the stationary distribution $\vec{q}$ of $L$-grams to get:
\begin{equation}
\label{eqn:entropy_Lgram_Markov}
S^{(L)}(\vec{q})
=
S(\vec{\pi})
-
(L-1) \sum_{{\mu}{\nu}} W_{{\nu}{\mu}} \pi_{\mu} \log{W_{{\nu}{\mu}}}.
\end{equation}
Typical values for the Shannon entropy of  $L$-grams are shown in Fig.~\ref{fig:markov-Lgrams}B along with the convergence curve of the NSB estimator.
We assume that the $L$-grams of a second system are generated by a similar Markov process but with a transition matrix $\hat{V}$ and stationary probabilities of the $\vec{\sigma}=\lbrace\sigma_{\mu}\rbrace$ states.
The cross-entropy between the $\vec{t}$ and $\vec{q}$ distributions reads:
\begin{equation}
\label{eqn:crossentropy_Lgram_Markov}
H^{(L)}(\vec{q}\Vert\vec{t})
=
H(\vec{\pi}\Vert\vec{\sigma})
-
(L-1) \sum_{{\mu}{\nu}} W_{{\nu}{\mu}} \pi_{\mu} \log{V_{{\nu}{\mu}}}.
\end{equation}

Similarly to the analysis in the previous section, we generate a large sample of $L$-grams from each distribution, with $N=M=2\cdot10^8$.
We subsample this dataset at different sample sizes and estimate the {\DKL} and its standard deviation for $L=2,3,4$. 
To study the average behavior, we {\IN divide the estimate} by the expected result (Eq.~\ref{eqn:crossentropy_Lgram_Markov}) and we {\IN average over 30 simulations}. 

We observe that, in the case of small numbers of categories ($K=20^2$,
Fig.~\ref{fig:markov-Lgrams}C top panel), {\IN DPM (and DP) perform
  quite similar to the best alternative (Jeffreys), but with different
  sign biases.}  However, the DPM estimator performance greatly
improves for larger $K$ (Fig.~\ref{fig:markov-Lgrams}C middle and
bottom panels).  In all cases, the standard deviation associated to
the DPM estimator (red bars in Fig.~\ref{fig:markov-Lgrams}C) captures
the spread across the different repetitions of the convergence curve
(red shade in Fig.~\ref{fig:markov-Lgrams}C).

\subsection{ Estimator for the Hellinger divergence }

\begin{figure}
\begin{center}
\includegraphics[width=\columnwidth]{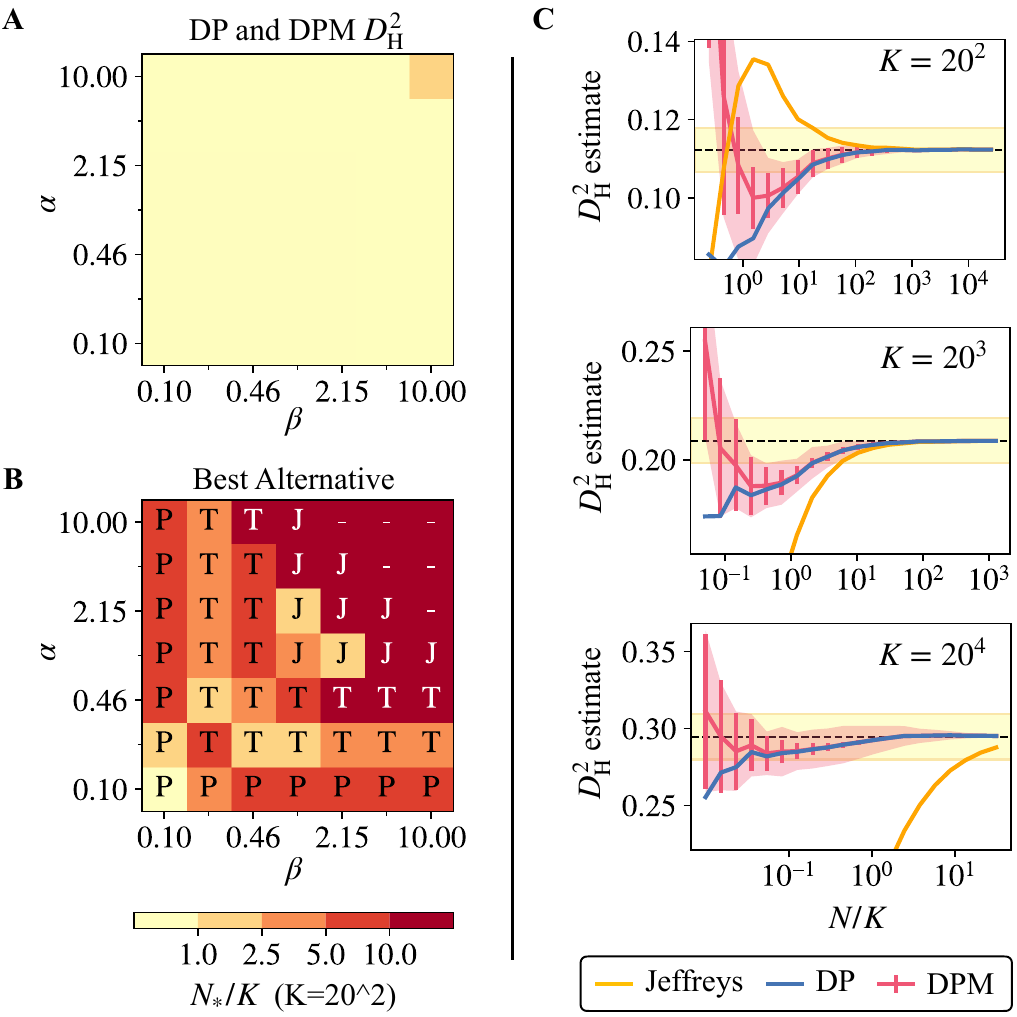}
\caption{
Squared Hellinger divergence convergence.
\textbf{A.}
Convergence score $N_{\ast}/K$ of the DP and DPM {$\Hell^2$}
estimators, tested on the same synthetic data as in Fig.~\ref{fig:dirichlet-convergence-estimation}B.
We consider 100 pairs of histograms drawn from Dirichlet-multinomial process with $\alpha$ and $\beta$ concentration parameters.
\textbf{B.} 
We compare to the score of the best alternative method to the DP and DPM, chosen as pseudo-count estimators with pseudo-count given by Table~\ref{tab:plugin-estimators}.
These alternative methods perform worse for all values of the parameters.
\textbf{C.} 
Convergence of the DP and DPM {$\Hell^2$} estimators tested on the same Markov datasets as in Fig.~\ref{fig:markov-Lgrams}C for different sub-sample sizes $N$.
Each repetition is normalized by its true value, averaged and then rescaled by the asymptotic value (average of the true values). 
Red shade: the standard deviation across repetitions.
}
\label{fig:test-for-hellinger}
\end{center}
\end{figure}

Lastly, we extend the DPM method to estimate the Hellinger divergence {\Hell} between the discrete distributions $\vec q$ and
$\vec t$~\cite{hellinger_neue_1909}.
The Hellinger divergence is a symmetric statistical distance that satisfies the triangular inequality, making it a true distance in the mathematical sense~\cite{liese_statistical_2008}:
\begin{equation}
\label{eqn:Hellinger-definition}
\Hell(\vec{q},\vec{t})^2
=
\frac{1}{2}\sum_{i=1}^{K}\left(\sqrt{q_{i}}-\sqrt{t_{i}}\right)^2=1-\sum_{i=1}^K\sqrt{q_it_i}.
\end{equation}
Following the same approach as for the Kullback-Leibler divergence (details in~{\apdx}~\ref{AppHell}), we obtain the DPM estimator for $\Hell^2$:
\begin{align}
\label{eqn:Hellinger-final-estimator}
\nonumber
&\langle \Hell^2 \vert \vec{n},\vec{m} \rangle
=\\
&=
\frac{1}{Z} 
\int \! {d\alpha} {d\beta} \;
\rho_{\rm H} (\alpha,\beta) P \left( \vec{n}\vert\alpha\right) P \left( \vec{m}\vert\beta\right)
\langle \Hell^2 \vert \vec{n}, \vec{m};\alpha,\beta\rangle,
\end{align}
with 
\begin{align}
\label{Hell_details}
\nonumber
&\langle \Hell^2 \vert \vec{n}, \vec{m};\alpha,\beta\rangle 
=\\
&=
1-\sum_{i=1}^{K} \frac{B(\frac{1}{2}, N + K\alpha)}{B(\frac{1}{2},n_{i} + \alpha)} \frac{B(\frac{1}{2}, M + K\beta)}{B(\frac{1}{2},m_i + \beta)},
\end{align}
where $Z = \int \! {d\alpha} {d\beta} \; \rho_{\rm H} (\alpha,\beta) P \left( \vec{n}\vert\alpha\right) P \left( \vec{m}\vert\beta\right)$ and $B(x_1, x_2) = \Gamma(x_1)\Gamma(x_1)/ \Gamma(x_1+x_2)$ is the two-dimensional Beta function.

We test the Hellinger divergence {\Hell} estimator on the same synthetic datasets as in Fig.~\ref{fig:dirichlet-convergence-estimation}B (Fig.~\ref{fig:test-for-hellinger}).
For datasets generated with Dirichlet-multinomial distributed samples, the DPM outperforms all considered plugin estimators
$1 - \sum_{i=1}^K \sqrt{\hat q_i\hat t_i}$, with $\hat q_i$ and $\hat t_i$ defined as before with pseudo-counts
$a,b$ chosen according to Table~\ref{tab:plugin-estimators} (Fig.~\ref{fig:test-for-hellinger}A).
As for the case of KL divergence, the performance improves for larger categories (Fig.~\ref{fig-si:dh-dirichlet-larger-K}).
Tests on the synthetic Markovian $L$-grams (see previous paragraph) show the DPM estimator performs better for larger numbers of categories $K$, with comparable performance to the best alternative (Jeffreys) for $K=20^2$ (Fig.~\ref{fig:test-for-hellinger}B).

\section{Discussion}
Correctly estimating statistical divergences between two distributions is an open problem in the analysis of categorical systems.  Alongside the entropy, divergences such as the Kullback-Leibler and the Hellinger distance, are an important tool in the analysis of categorical data~\cite{zhang_entropic_2022}.

We focused on categorical distributions with finite numbers of categories $K$ (bounded domain), where $K$ is a known quantity.  We proposed a  way (DPM) to extend the approach of Nemenman et al.~\cite{nemenman_entropy_2001} developed for Shannon entropy estimation, to Kullback-Leibler estimation. DPM introduces a 
mixture of symmetric Dirichlet priors with a log-uniform  \textit{a priori} expected divergence distribution (Eq.~\ref{eqn:meta_prior-phi_func}).  
We restricted our analysis to the case of the two finite samples of the same size $N$, although the method works for different sample sizes.
We also propose a simplified estimator (DP), which assumes a Dirichlet prior with concentration parameter fixed to the maximum value of the likelihood.
This estimator is faster to compute as it does not require to integrate over the concentration parameters.

We showed that the DPM method outperforms the tested empirical plugin techniques 
in terms of  {\DKL} estimation for synthetic data sampled from  a Dirichlet-multinomial distribution with fixed  concentration parameters. The estimation task gets harder for distributions with larger concentration parameters, {\IN i.~e.}, closer to a uniform distribution, but easier for large numbers of categories $K$. 

These convergence trends were confirmed by tests on sequences of $L$ states generated by Markov chains.
In this case, DPM compares well to the best plugin estimator in the low sample size regime of $K=20^2$ and outperforms it for $K\geq 20^3$.
Similar results were obtained for the  DPM estimate of the Hellinger divergence for both Dirichlet-multinomial and Markov chain datasets.
{\IN 
To our knowledge, DPM estimator of the Hellinger divergence is the first attempt to extend the ideas of Ref.~\cite{nemenman_entropy_2001} and to build a uniform prior estimator for a non-entropy-like quantity.
}

Our tests were restricted to categorical systems with rank distributions having exponentially decaying tails. 
As previously discussed for the case of the NSB entropy estimator, the Dirichlet prior has major limitations in capturing the Shannon entropy if the  system rank distribution is not decaying fast~\cite{archer_bayesian_2014, hernandez_low-probability_2023}.
Many real systems exhibit long-tailed rank distributions that decay as power-laws~\cite{zipf_human_1949}, which are not well captured by a Dirichlet prior.
Preliminary (unpublished) tests of the DPM method for such systems show poor performance.
Similarly to the case of entropy estimates, we speculate that the limitations of this method are related to issues with the poor representation of long tails by Dirichlet priors. 
Introducing a Pitman-Yor prior~\cite{pitman_two_parameter_1997} could overcome this problem, as has been shown for entropy estimation by Archer et al.~\cite{archer_bayesian_2014}, and offers a direction to generalize the applicability of the DPM method.
Extending the Pitman-Yor prior to the case of statistical divergences would require to compute expected values over the probabilities of both systems, but to the best of our knowledge this is not possible because of the lack of an analytical expression for the Pitman-Yor distribution.
Another difficulty may lie in the difficulty to encode correlations between the ranks of categories in the two distributions.
Our priors assume that the two unknown distributions $\vec q$ and $\vec t$ are drawn independently. However in real data they are generally correlated, which could have an impact on the quality of estimators when the distribution of frequencies becomes very broad.

In view of these complications, it is important to have practical criteria to ascertain if the output of the DPM estimator can be trusted for a specific dataset, or if it remains biased. Similar questions exist for estimation of many quantities, and specifically of entropic quantities, on categorial data since no estimator can be universally unbiased for them, and the decay of the bias with the sample size may be excruciatingly slow \cite{antos_convergence_2001,paninski_estimation_2003}. For entropy and mutual information, the standard approach is to observe if the empirical output drifts systematically as the sample size changes. One then declares the estimator trustworthy if the bias does not drift by more than the posterior standard deviation over about an order of magnitude change in the amount of data \cite{strong_entropy_1998,holmes2019estimation}. We expect this approach to transfer nearly verbatim to the \DKL~and the Hellinger divergence context, easily detecting whether the  DPM approach can be used for a specific dataset, or if other analysis methods should be sought.

\section*{Acknowledgements}
{
We thank Antonio C.\ Costa for helpful discussions.
This work was partially supported by the
European Research Council Consolidator Grant n.~724208 and ANR-19-CE45-0018 ``RESP-REP'' from the Agence Nationale de la Recherche.
I.N.\ was supported in part by the Simons Foundation Investigator grant and by the U.~S.\ NSF grant No.~2209996.
}

\bibliographystyle{unsrt}

\appendix*
\section{supplementary information}
\setcounter{equation}{0}
\renewcommand{\theequation}{A.\arabic{equation}}
\subsection{ Mathematical relations }
\label{AppDirMult}

We first introduce mathematical relations and notations that are used for the computation of the DP and DPM estimators for $\DKL$ and ${\Hell}^2$.

\subsubsection{ Wolpert-Wolf integrals }
Given a vector $\vec{x}=\lbrace x_i \rbrace_{i=1}^K$, where $x_i \in (0,\infty)$ for all  $i=1,\dots, K$, where $K$ is a finite number of categories, the Wolpert-Wolf \cite{wolpert_estimating_1995} integral is  a multivariate Beta function $B:\mathbb{R}_{+}^K\rightarrow\mathbb{R}_{+}$ in $\vec{x}$:
\begin{equation}
\label{eqn:multivare-beta-wolpert-wolf}
\int \! {d\vec{q}} \; \delta\left( \sum_{i=1}^K q_{i}-1\right)  \prod_{j=1}^K {q_j} ^{x_j-1} 
= \dfrac{\prod_{j=1}^K \Gamma (x_j)}{\Gamma(X)} = B(\vec{x}),
\end{equation}
where  $X=\sum_{i} x_i$ and the function $\Gamma$ is the Euler Gamma function $\Gamma(x) = \int_0^{\infty} \! {dt} \; e^{-t}t^{x-1}$. 
All Bayesian calculations with multinomial likelihoods and multivariate Dirichlet priors involve the integral:
\begin{align}
\nonumber
&\int \! {d\vec{q}} \; \delta\left( \sum_{i=1}^K q_{i}-1\right) \prod_{j=1}^K f_j(q_j)
=\\
&=
{\cal L}^{-1} \left[ \prod_{j=1}^K {\cal L} \left[ f_j(q) \right](s)\right](q'=1),
\end{align}
where the $f_i$ are regular functions,  ${\cal L}$ is the Laplace transform in $q$ (which is a function of $s$)  and ${\cal L}^{-1}$ is the inverse Laplace transform (which is a function of $q'$).

\subsubsection{ Partial derivative operation }

The ``partial derivative operator'' for the $i$-th dimension $\partial_i = \frac{\partial}{\partial x_{i}}$  applied to the Beta function $B$ returns 
\begin{align}
\label{eqn:WW-derivation-rule}
\nonumber
&
\left( \partial_{i} B \right) (\vec{x} )
=
\int \! {d\vec{q}} \; \delta(\Vert\vec{q}\Vert_1-1)  \prod_{j=1}^K {q_j} ^{x_j -1} \log q_i
\\
&= B\left(\vec{x}\right) \left[ \psi(x_i)-\psi(X) \right],
\end{align}
where the function $\psi$ is the polygamma function of order $0$. The polygamma function of order $\ell$ is defined as 

\begin{equation}
\label{eqn:polygamma}
\psi_{\ell}(x) = \left.\frac{d^{\ell} }{d y^{\ell}}\log \Gamma(y)\right|_{y=x}.
\end{equation}

In order to simplify the calculations, we define the following quantities related to the partial derivative operation (Eq.~\ref{eqn:WW-derivation-rule}):
\begin{equation}
\Lambda_i\left(\vec{x}\right)
= 
\frac{ \left( \partial_{i} B \right) (\vec{x} )}{B\left(\vec{x}\right)}
=
\Delta \psi(x_i,X) ,
\label{eqn:WW-simple-deriv}
\end{equation}
where we make use of the contraction  $\Delta \psi(z_1,z_2) = \psi(z_1)-\psi(z_2)$. Iterating this derivation on the function $B$, one can express the double derivatives as follows:
\begin{align}
\label{eqn:WW-double-deriv}
\nonumber
&
\Lambda_{ij}\left(\vec{x}\right)
=
\frac{ (\partial_{ij} B) \left( \vec{x}\right)}{B\left( \vec{x}\right)}
=
\frac{ \left(\partial_{i} \left( B \Lambda_{j} \right)\right)\left( \vec{x}\right)}{B\left( \vec{x}\right)}
\\
&=
 \Lambda_{i}\left( \vec{x}\right)\Lambda_{j}\left( \vec{x}\right) + \left( \partial_{i} \Lambda_{j} \right) \left( \vec{x}\right)
,
\end{align}
where the derivative $\partial_{i} \Lambda_{j}\left( \vec{x}\right) = \delta_{ij}\psi_{1}\left( x_i\right) -\psi_{1}\left( X\right)$ is a consequence of Eq.~\ref{eqn:WW-simple-deriv}.

\subsubsection{ Shift Operation }

The ``shift operator'' $e^{\lambda\partial_i}$ of parameter $\lambda \in \mathbb{R}$ for the $i$-th dimension acts on the function $B$ as follows:
\begin{align}
\label{eqn:WW-translation-rule}
\nonumber
&\left( e^{\lambda\partial_i}B \right)\left( \vec{x}\right)
=
B\left( \vec{x}+\lambda \widehat{i}\right) 
\\
&=
\int \! {d\vec{q}} \; \delta\left( \sum_{i=1}^K q_{i}-1\right) \prod_{j=1}^K {q_j} ^{x_j-1} {q_i}^{\lambda}
=
B\left( \vec{x}\right) \dfrac{B \left( \lambda , X\right)}{B\left(\lambda, x_{i}\right) }
,
\end{align}
where $\widehat{i} = \lbrace\delta_{ij}\rbrace_{j=1}^K$ indicates the $i$-th versor in the $K$-dimensional space of categories, with the condition $x_{i}+\lambda>0$. The function $B\left( z_{1},z_{2}\right)$ is the regular (two-dimensional) Beta function :
\begin{equation}
\label{eqn:WW-beta-function}
B\left( z_{1},z_{2}\right)
=
\dfrac{\Gamma \left( z_{1}\right) \Gamma \left( z_{2}\right) }{\Gamma \left( z_{1}+z_{2}\right) }
.
\end{equation}

When $\lambda =n \in \mathbb{N}_{+}$, the shift simplifies to
\begin{equation}
\label{eqn:WW-positive-int-shift-rule}
\left( e^{n\partial_i} B \right) \left( \vec{x}\right)
=
B\left( \vec{x}\right) \prod ^{n-1}_{n'=0}\dfrac{x_{i}+n'}{X+n'}
\end{equation}
as an immediate consequence of the recurrence relation $\Gamma (x + 1) = x \Gamma(x)$.  
Similarly to the case of  partial derivatives, we introduce a class of functions to deal with the shift:
\begin{equation}
\label{eqn:WW-omega-simple-shift}
\Omega _{i} \left( \vec{x}\right) 
=
\frac{\left(e^{\partial_i}B \right)\left( \vec{x}\right)}{B \left( \vec{x}\right)}
=
\dfrac{x_{i}}{X} 
,
\end{equation}
from which we compute two-dimensional shifts
\begin{align}
\label{eqn:WW-omega-double-shift}
\nonumber
&\Omega _{ij} \left( \vec{x}\right)
=
\frac{ \left( e^{\partial_i}e^{\partial_j}B \right) \left( \vec{x}\right)}{B\left( \vec{x}\right)}
=
\frac{ e^{\partial_i} \left( B \Omega_{j} \right) \left( \vec{x}\right)}{B\left( \vec{x}\right)}
\\
&=
\Omega_{i}\left( \vec{x}\right) \Omega_{j}\left( \vec{x} + \widehat{i}\right)
=
\dfrac{x_{i}}{X} \frac{( x_{j} +\delta _{ij}) }{X+1}.
\end{align}

\subsubsection{ Composed operations }

For the sake of this work, another useful class of functions are the first order derivatives of the functions $\Omega$ defined as
\begin{equation}
\label{eqn:WW-omega-shift_1-deriv_1}
\frac{\partial_{j} \Omega_{i}}{\Omega_{i} }
= 
\partial_{j} \log\Omega_{i}
=
\frac{\delta_{ij}}{x_j} - \frac{1}{X}
,
\end{equation}
and, for the two-dimensional shift,
\begin{align}
\nonumber
\label{eqn:WW-omega-shift_2-deriv_1}
&\frac{\partial_{k} \Omega_{ij}}{\Omega_{ij}}
=
\partial_{k} \log \Omega_{ij}
\\
\nonumber
&=
\delta_{ijk} \left( \frac{1}{x_k} + \frac{1}{x_k+1} \right)
+ (\delta_{ik} + \delta_{jk} ) (1-\delta_{ij})\frac{1}{x_k}
+ \\
&  \hphantom{=}
- \frac{1}{X} - \frac{1}{X+1}
.
\end{align}

Similarly for the second order derivatives:
\begin{align}
\label{eqn:WW-omega-shift_1-deriv_2}
\nonumber
&
\frac{\partial_{jk} \Omega_{i}}{\Omega_{i} }
=
\partial_{j}\partial_{k} \log \Omega_{i} + \left(\partial_{j} \log \Omega_{i}\right) \left(\partial_{k} \log \Omega_{i}\right)
\\
&=
\frac{2}{X^2} - \frac{\delta_{ij}+\delta_{ik}}{x_i  X}
,
\end{align}
and 
\begin{align}
\label{eqn:WW-omega-shift_2-deriv_2}
\nonumber
&
\frac{\partial_{kh} \Omega_{ij}}{\Omega_{ij}}
=
\frac{\partial_{k} (\Omega_{ij} \partial_{h}\log \Omega_{ij})}{\Omega_{ij}}
\\
\nonumber
&=
\frac{1}{X^2} + \frac{1}{(X+1)^2}
- \delta_{ijkh} \left( \frac{1}{{x_k}^2} + \frac{1}{(x_k+1)^2} \right)
+ \\
&  \hphantom{=}
-(\delta_{ikh} + \delta_{jkh} ) (1-\delta_{ij})\frac{1}{x_h^2}
+ \left(\partial_{k} \log \Omega_{ij}\right) \left(\partial_{h} \log \Omega_{ij}\right).
\end{align}

Using all these definitions, we compute the following quantities:

\begin{equation}
\label{eqn:WW-deriv_1+shift_1}
\frac{ e^{\partial_i} \partial _{j} B}{B}
= 
\Omega_{i} \Lambda_j + \partial_{j} \Omega_{i},
\end{equation}

\begin{equation}
\label{eqn:WW-deriv_1+shift_2}
\frac{ e^{\partial_i} e^{\partial_j}\partial _{k} B}{B}
= 
\Omega_{ij} \Lambda_k + \partial_{k} \Omega_{ij},
\end{equation}
\begin{equation}
\label{eqn:WW-deriv_2+shift_1}
\frac{e^{\partial_i} \partial _{jk} B}{B}
=
\Omega_i \Lambda_{jk} + (\partial_j \Omega_i) \Lambda_k
+
(\partial_k \Omega_i) \Lambda_j + \partial_{jk} \Omega_i
\end{equation}
and
\begin{equation}
\label{eqn:WW-deriv_2+shift_2}
\frac{e^{\partial_i} e^{\partial_j}\partial _{kh} B}{B}
= 
\Omega_{ij} \Lambda_{kh} + (\partial_k \Omega_{ij}) \Lambda_h
+
(\partial_h \Omega_{ij}) \Lambda_k + \partial_{kh} \Omega_{ij}.
\end{equation}
which are used to reconstruct all estimators presented in this work.

\subsubsection{A priori and a posteriori expected values }

The operations presented in the previous sections are used compute the posterior expected values $\langle F(\vec{q},\vec{t}) \vert \vec{n},\vec{m};\alpha, \beta \rangle$ for all the functions that can be expressed as :
\begin{equation}
\label{eqn:computability-condition}
F(\vec{q},\vec{t})
 =
 \sum_{i=1}^K f_i(\vec{q}) g_i (\vec{t}).
\end{equation}
Since the concentration parameters $\alpha$, $\beta$ are independent,  for fixed concentration parameters the expected value of $F$ factorizes:
\begin{equation}
\langle F(\vec{q},\vec{t}) \vert \vec{n},\vec{m};\alpha, \beta \rangle
=
\sum_{i=1}^K \langle f_i \vert \vec{n};\alpha\rangle  \langle g_i \vert \vec{m};\beta \rangle,
\end{equation}
with
\begin{align}
\nonumber
&
\langle f_i \vert \vec{n};\alpha\rangle \frac{B(\vec{n}+\vec{\alpha})}{B(\vec{\alpha}) B(\vec{n}+\vec{1})}
\\
\nonumber
&=
\int \! {d \vec{q}} \;
\delta\left( \sum_{i=1}^K q_{i}-1\right)
\Dir{q}{\alpha} \Mult{n}{q}
f_i(\vec{q}) 
\\
&=
\int \! {d \vec{q}} \;
\delta\left( \sum_{i=1}^K q_{i}-1\right)
\frac{\prod_{j}^K {q_j} ^{n_i+\alpha-1}}{ B(\vec{\alpha}) B(\vec{n}+\vec{1})}
f_i (\vec{q}).
\end{align}
{\IN
For all functions $f_i$ that can be expressed in terms of partial derivative (Eq.~\ref{eqn:WW-derivation-rule}) and/or shift operators (Eq.~\ref{eqn:WW-translation-rule}),
a factor $B(\vec{n}+\vec{\alpha})$ appears  and the expected value is obtained explicitly simplifying the constant factors.
}
Specifically:
\begin{equation}
\label{eqn:posterior-q_i}
\langle q_i \vert \vec{n};\alpha\rangle
=
\frac{ \left(e^{\partial_{i}} B \right)(\vec{n}+\vec{\alpha})}{B(\vec{n}+\vec{\alpha})},
\end{equation}
\begin{equation}
\label{eqn:posterior-log-q_i}
\langle \log q_i \vert \vec{n};\alpha\rangle
=
\frac{ \left(\partial_{i} B \right)(\vec{n}+\vec{\alpha})}{B(\vec{n}+\vec{\alpha})},
\end{equation}
\begin{equation}
\label{eqn:posterior-q_i-log-q_i}
\langle q_i \log q_i \vert \vec{n};\alpha\rangle
=
\frac{ \left(e^{\partial_{i}}  \partial_{i} B \right)(\vec{n}+\vec{\alpha})}{B(\vec{n}+\vec{\alpha})},
\end{equation}
\begin{equation}
\label{eqn:posterior-q_i-q_j}
\langle q_i q_j \vert \vec{n};\alpha\rangle
=
\frac{ \left(e^{\partial_{i}}  e^{\partial_{j}} B \right)(\vec{n}+\vec{\alpha})}{B(\vec{n}+\vec{\alpha})},
\end{equation}
\begin{equation}
\label{eqn:posterior-q_i-q_j-log-q_i}
\langle q_i q_j \log q_i \vert \vec{n};\alpha\rangle
=
\frac{ \left(e^{\partial_{i}}  e^{\partial_{j}}  \partial_i B \right)(\vec{n}+\vec{\alpha})}{B(\vec{n}+\vec{\alpha})}
\end{equation}
and
\begin{equation}
\label{eqn:posterior-q_i-q_j-log-q_i-log-q_j}
\langle q_i q_j \log q_i \log q_j \vert \vec{n};\alpha\rangle
=
\frac{ \left(e^{\partial_{i}}  e^{\partial_{j}} \partial_i \partial_j B \right)\left(\vec{n}+ \vec{\alpha} \right) }{B(\vec{n}+\vec{\alpha})}
.
\end{equation}

The \textit{a priori} expected values are computed in the same way, noticing that $\langle f_j \vert \alpha\rangle = \langle f_j \vert \vec{n}=\vec{0};\alpha\rangle$. 

\subsubsection{KL divergence estimation}

We can use the previous results to compute the \textit{a posteriori} expected value for the {\DKL}.
We start by computing the \textit{a posteriori} expected value for the crossentropy $H$ which is given by
\begin{align}
\nonumber
&
\left\langle
\sum_{i}^{K} 
H(\bm{q} \Vert \bm{t})
\vert
\bm{n},\bm{m}, \alpha, \beta
\right\rangle
=
 \sum_{i}^{K} 
 \langle q_i \vert \bm{n}, \alpha \rangle 
 \langle \log t_i \vert \bm{m}, \beta \rangle
 \\
 \nonumber
 &=
  \sum_{i}^{K} 
  \frac{ e^{\partial_i} B(\bm{n}+\bm{\alpha}) }{B(\bm{n}+\bm{\alpha})}
  \frac{ \partial_i B(\bm{m}+\bm{\beta}) }{B(\bm{m}+\bm{\beta})}
 \\
&=
  \sum_{i}^{K}  \frac{n_i+\alpha}{N+K\alpha} \Delta \psi (M+K\beta, m_i+\beta)
  ,
 \end{align}
where we took advantage of independence and used the relations Eq.~\ref{eqn:posterior-q_i} and \ref{eqn:posterior-log-q_i} to obtain the explicit expressions in Eq.~\ref{eqn:WW-omega-simple-shift} and \ref{eqn:WW-simple-deriv}.
Subtracting the \textit{a posteriori} expected Shannon entropy $\langle S \vert \bm{n},\bm{m}, \alpha, \beta
\rangle = \sum_i \dfrac{n_i+\alpha}{N+K\alpha} \Delta \psi(N+K\alpha+1, n_i +\alpha+1)$, we finally obtain the KL expected value in Eq.~\ref{eqn: posterior-DKL}:
\begin{align}
\nonumber
&\langle \DKL \vert \vec{n},\vec{m} ; \alpha,\beta\rangle 
\\
\nonumber
&=
\sum_i\frac{n_i+\alpha}{N+K\alpha} 
\lbrace \Delta\psi(M+K\beta , m_i+\beta) 
\\
&\hphantom{=}
- \Delta\psi(N+K\alpha+1 , n_i+\alpha+1) \rbrace.
\end{align}

\subsubsection{Squared KL divergence estimation}

In order to compute the {\IN posterior} standard deviation of the Kullback-Leibler divergence estimator, we calculate the expected value of the squared KL divergence:
\begin{align}
\label{eqn:squared-DKL-estimator}
\nonumber
&\langle {\DKL}^2 \vert \vec{n},\vec{m} \rangle
\\
&=
\int \! {d\alpha} {d\beta} \;
P(\vec{n},\alpha)P(\vec{m}\vert,\beta)\rho(\alpha,\beta)
\langle {\DKL}^2 \vert \vec{n},\vec{m};\alpha,\beta \rangle
.
\end{align}

Similarly to the case of $\DKL$,
{\IN
we can compute explicitly 
}
\begin{align}
\langle {\DKL}^2 \vert \vec{n},\vec{m};\alpha,\beta \rangle
=
\sum_{ij} \langle q_i q_j \log\frac{q_i}{t_i}\log\frac{q_j}{t_j} \vert \vec{n},\vec{m};\alpha,\beta \rangle
,
\end{align}
{\IN which requires to rewrite}
\begin{align}
\label{eqn:squared-DKL-definition}
\nonumber
&
q_{i}q_{j}\log \dfrac{q_{i}}{t_{i}}\log \dfrac{q_{j}}{t_{j}}
\\
&=
q_{i}q_{j}\log q_{i}\log q_{j} 
- 2 q_{i}q_j\log q_i\log t_{j}
+ q_{i}q_{j}\log t_i\log t_{j}
. 
\end{align}

The explicit expression computed using Wolpert-Wolf properties (Eqs.~\ref{eqn:WW-simple-deriv},~\ref{eqn:WW-double-deriv}, ~\ref{eqn:WW-omega-double-shift}, ~\ref{eqn:WW-deriv_1+shift_2}, ~\ref{eqn:WW-deriv_2+shift_2}) is:
\begin{align}
\label{eqn:squared-DKL-ww-expression-explicit-pt2}
\nonumber
&
\langle q_i q_j \log\frac{q_i}{t_i}\log\frac{q_j}{t_j} \vert \vec{n},\vec{m};\alpha,\beta \rangle
\\
\nonumber
&=
\frac{x_{i}(x_{j}+\delta_{ij})}{X(X+1)}
\Big\lbrace 
 \delta_{ij}\psi_{1}\left( x_i+2 \right) -\psi_{1} \left( X+2 \right)
\\
\nonumber
& \hphantom{=}
+ \Delta \psi\left( x_{i}+1+ \delta_{ij},X+2\right) \cdot (i\leftrightarrow j) 
\\
\nonumber
&  \hphantom{=}
- 2\Delta \psi \left( x_{i}+1+\delta_{ij},X+2\right)
\Delta \psi \left( y_{j},Y\right)
\\
& \hphantom{=}
+\delta_{ij} \psi _{1}(y_{i}) - \psi _{1} (Y)
+
\Delta \psi \left( y_{i},Y\right)\cdot\left(i\leftrightarrow j\right)
\Big\rbrace
,
\end{align}
where we have introduced the following notation to contract the expression: $\vec{x}=\vec{n}+\vec{\alpha}$, $X=N+K\alpha$, $\vec{y}=\vec{m}+\vec{\beta}$ and $Y=M+K\beta$.
The factor $\left(i\leftrightarrow j\right)$ means taking the term that it multiplies with inverted indexes $i$ and $j$.

\subsection{Zhang-Grabchak divergence estimator}
\label{AppZhang}

In Ref.~\cite{zhang_nonparametric_2014} Zhang and Grabchak proposed an estimator for the Kullback-Leibler divergence, defined as:
\begin{align}
\label{eqn:z-estimator-original}
\nonumber
\widehat{D}_{\rm KL}^{(z)}
=
\sum_{i=1}^K \frac{n_i}{N}
\Biggl\lbrace
&
\sum_{v=1}^{M-m_i}\frac{1}{v}\prod_{s=1}^{v}\left(1 - \frac{m_i}{M-s+1} \right)
\\
&-
\sum_{v=1}^{N-n_i}\frac{1}{v}\prod_{s=1}^{v}\left(1 - \frac{n_i-1}{N-s} \right)
\Biggr\rbrace
,
\end{align}
where $v$ and $s$ are dummy variables. 

\subsubsection{Expression of the Z-estimator}

Schurmann \cite{schurmann_note_2015} has shown that, in the entropy term of Eq.~\ref{eqn:z-estimator-original}, the summation in $v$ of the $i$-th element can actually be expressed in a more concise way as 
\begin{equation}
\label{eqn:z-shurmann-resummation}
\sum_{v=1}^{N-n_i}\frac{1}{v}\prod_{s=1}^{v}\left(1 - \frac{n_i-1}{N-s} \right)
=
\Delta \psi (N, n_{i})
,
\end{equation}
times a factor $n_{i}/N$.
The sum of these terms returns the Shurman-Grassberger entropy estimator $\widehat{S}_{\rm SG} = \sum_{i=1} \frac{n_i}{N} \Delta \psi (N, n_{i})$ \cite{schurmann_bias_2004}.

If we simply plug $N=M+1$ and $n_{i} = m_{i}+1$ in Eq.~\ref{eqn:z-shurmann-resummation}, we can show that the analogous $i$-th crossentropy term in Eq.~\ref{eqn:z-estimator-original} is equal to the following:
\begin{equation}
\label{eqn:z-shurmann resummation}
\sum_{v=1}^{M-m_i}\frac{1}{v}\prod_{s=1}^{v}\left(1 - \frac{m_i}{M-s+1} \right)
=
\Delta \psi (M+1, m_{i}+1)
.
\end{equation}

Finally, if we substitute Eq.~\ref{eqn:z-shurmann-resummation} and \ref{eqn:z-shurmann resummation} in Eq.~\ref{eqn:z-estimator-original}, we obtain:
\begin{equation}
\label{eqn:z-estimator-final-apdx}
\widehat{D}_{\rm KL}^{(z)}
=
\sum_{i=1}^K \frac{n_i}{N}
\left[
\Delta \psi (M+1, m_{i}+1)-\Delta \psi (N, n_{i})
\right]
.
\end{equation}
which is the expression in Eq.~\ref{eqn:z-estimator-final} of the main text.

\subsubsection{ Relation between the DP and the Z estimator }

The Z-estimator can be expressed as an \textit{a posteriori} expected value of the {\DKL} at $\alpha=0$ and $\beta=1$, up to an additive constant.
We start by showing the following relation
\begin{align}
\label{eq:aposteriori-S-to-Z}
\nonumber
&
\lim_{\alpha\to 0}
\langle S \vert \bm{n}, \alpha \rangle
= 
\sum_{i=1}^K \frac{n_i}{N}\Delta \psi (N+1, n_{i}+1)
\\
&=
\frac{1-K}{N}
+ 
\sum_{i=1}^K \frac{n_i}{N} \Delta \psi (N, n_{i})
,
\end{align}
which makes use of the fact that $\psi(x+1)=\psi(x)+\frac{1}{x}$.

Considering now the crossentropy term with $\beta=1$, and performing the same limit as before, we observe that
\begin{align}
\label{eq:aposteriori-H-to-Z}
\nonumber
&
\lim_{\alpha\to 0}
\langle H \vert \bm{n}, \bm{m}, \alpha, \beta
=1 \rangle
=
\sum_{i=1}^K \frac{n_i}{N}\Delta \psi (M+K, m_{i}+1)
\\
&=
\Delta \psi (M+K, M+1) +\sum_{i=1}^K \frac{n_i}{N}\Delta \psi (M+1, m_{i}+1)
,
\end{align}
where we used the fact that $\Delta \psi(x, x) = \psi(x)-\psi(x)=0$ to add the term $\Delta \psi (M+1,M+1)$ in the sum.

Recognizing the two terms in Eq.~\ref{eqn:z-estimator-final-apdx} we subtract Eq.~\ref{eq:aposteriori-S-to-Z} and \ref{eq:aposteriori-H-to-Z} to obtain that 
\begin{align}
\label{eq:aposteriori-DKL-to-Z}
\nonumber
&
\lim_{\alpha\to 0}
\langle \DKL \vert \bm{n}, \bm{m}, \alpha, \beta=1 \rangle
\\
&=
\Delta \psi (M+K, M+1)
+
\frac{K-1}{N}
+
\widehat{D}_{\rm KL}^{(z)}
.
\end{align}

\subsection{The DPM squared Hellinger divergence estimator }
\label{AppHell}

We compute the DPM estimator for the squared Hellinger divergence $\Hell^2$ (Eq.~\ref{eqn:Hellinger-definition}).
We do so by starting from the Bhattacharyya coefficient {\Bhatt} \cite{bhattacharyya_on_1943}
\begin{equation}
{\Bhatt}(\vec{q},\vec{t})
=
\sum_{i=1}^{K}\sqrt{q_{i}}\sqrt{t_{i}}
=
1 - {\Hell}^2(\vec{q},\vec{t})
.
\end{equation}
Its \textit{a priori} expected value under the assumption of the prior 
\begin{equation}
\label{eqn:vanilla-Dirichlet-prior}
P_{\rm prior}(\vec{q},\vec{t}) =p( \vec{q},\vec{t} \vert \alpha,\beta)
=
\Dir{q}{\alpha} \Dir{t}{\beta}
\end{equation} 
is equal to :
\begin{align}
\nonumber
&
\langle {\Bhatt} \vert \alpha , \beta \rangle
=
\sum_{i=1}^{K} \frac{(e^{\frac{1}{2}\partial_{i}}B)(\vec{\alpha})}{B(\vec{\alpha})} \frac{(e^{\frac{1}{2}\partial_{i}}B)(\vec{\beta})}{B(\vec{\beta})}
\\
&=
K \frac{B(\frac{1}{2}, K\alpha)}{B(\frac{1}{2},\alpha)} \frac{B(\frac{1}{2}, K\beta)}{B(\frac{1}{2},\beta)}
,
\end{align}
where we used the shift property (Eq.~\ref{eqn:WW-translation-rule}) with parameter $\lambda=\frac{1}{2}$. Following the derivation of the {\DKL} in the main text, we choose a metaprior $\rho_{\rm H} (\alpha,\beta)$ to control the \textit{a priori} squared Hellinger divergence $\langle {\Hell}^2 \vert \alpha , \beta \rangle = 1 - \langle {\Bhatt} \vert \alpha , \beta \rangle$:
\begin{equation}
\label{eqn:general-marginalization}
\rho_{\rm H}  (z) =
\int \! {d\alpha} {d\beta} \;
\rho_{\rm H} (\alpha,\beta) \, \delta\left( \langle \Hell^2 \vert \alpha , \beta \rangle -z \right).
\end{equation}

We define $g(x)= \sqrt{K} B(\frac{1}{2},  Kx) / B(\frac{1}{2},x)$, which is a function $g : \mathbb{R} \to [0,1)$.
Using a similar Ansatz of the one in the main text, we obtain $\rho_{\rm H} (\alpha,\beta) = \vert \partial_{\alpha} g (\alpha) \vert  \vert \partial_{\beta} g (\beta) \vert \phi( \langle \Hell^2 \vert \alpha , \beta \rangle )$, where the condition in Eq.~\ref{eqn:general-marginalization}  imposes 
\begin{equation}
\phi ( z ) = \rho_{\rm H} (z) \frac{(1-z)^2}{z(2-z)}.
\end{equation}
We choose $\rho_{\rm H} (z)$ to be log-uniform. 

Finally, knowing that the calculation for the posterior expected squared Hellinger divergence is analogous to the \textit{a priori} expectation, we obtain the DPM squared Hellinger estimator in Eq.~\ref{eqn:Hellinger-final-estimator} and~\ref{Hell_details}.

\subsection{Numerical implementation}
\label{AppNumImp}

\subsubsection{Computations with multiplicities}

In the low sampling regime (sparse data), there is a limited number of values the counts can take, which means that many categories will see the same pairs of values $\vec{x_i}=(n_i,m_i)$.
To reduce the computational cost associated to summation over the $K$ categories, we introduce a set of ``multiplicities'' \cite{paninski_estimation_2003} contained in the vector $\nu_{\vec{x}}$, where each entry is the number of instances that appear $n$ times in the first sample and $m$ in the second.
Since by construction the dimension of $\nu_{\vec{x}} \leq K$, we expressed all summation in terms of the multiplicities vector. Given a function of the two counts $f$,  the sum over all categories is:
\begin{equation}
\sum_{i=1}^K f(\vec{x_i}) = \sum_{ \vec{x} } \nu_{\vec{x}} f(\vec{x}),
\end{equation}
where the last sum runs over the ensemble of distinct pairs of observed counts. In the case of double sums (e.g. for $\DKL^2$), one needs to re-express the function as:
\begin{equation}
\begin{aligned}
f(\vec{x_i}, \vec{x_j}) 
=
\delta_{ij} f_{\parallel}(\vec{x_i}) + (1-\delta_{ij})f_{\bot}(\vec{x_i},\vec{x_j}),
\end{aligned}
\end{equation}
where $f_{\parallel}$ and $f_{\bot}$ is the function $f$ for $i=j$ and $i\neq j$.
The summation over the terms in $\delta_{ij}$ is calculated as before, and the double summation is
\begin{equation}
\begin{aligned}
\sum_{i,j} f_{\bot}(\vec{x_i}, \vec{x_j}) 
=
\sum_{ \vec{x},  \vec{x'}} \nu_{\vec{x}} \nu_{\vec{x'}}  f_{\bot}(\vec{x}, \vec{x'}).
\end{aligned}
\end{equation}
These formulas allow us to exploit vectorial expressions in the numerical calculations.

\subsubsection{Numerical integrations}

Similarly to Ref.~\cite{nemenman_coincidences_2011}, to compute numerically the quantities $\langle \DKL \vert \vec{n},\vec{m} \rangle$ (Eq.~\ref{eqn:DKL-estimator}) and $\langle {\DKL}^2 \vert \vec{n},\vec{m} \rangle$ (Eq.~\ref{eqn:squared-DKL-estimator}), we first seek for the maximum $(\alpha_{*},\beta_{*})$ of the quantity $\mathcal{L}(\alpha,\beta)$ (see Fig.~\ref{fig:evidence} for further details).
For accuracy, we perform this computation in logarithmic space of $\log \alpha$ and $\log \beta$.
Rescaling $\mathcal{L}(\alpha,\beta)$ by its maximum, integrands are $\mathcal{O}(1)$ for $(\alpha,\beta) \sim (\alpha_{*},\beta_{*})$.
To find the maximum of the log-evidence (minimum of the opposite), we use the ``Limited-memory BFGS'' optimization algorithm as coded in the function ``minimize'', module \textit{optimize} of the Python package \textit{scipy} (version 1.7.3).

We evaluate the integrals using the trapezoidal rule.  {\IN From the
  Hessian at the maximum of the log-evidence, we compute the standard
  deviation in the $\alpha$ and the $\beta$-direction as if the
  posterior was Gaussian.}  We use this standard deviation to pick a
range of parameters spanning 3 standard deviations on both sides of
the maximum.  {\IN We heuristically chose the number of bins within
  the ranges for the integration, to be equal to
  $10\left(\frac{K}{N}\right)^2$ for $\alpha$
  ($10\left(\frac{K}{M}\right)^2$ for $\beta$).  }

\subsection{Code availability}
\label{AppCodAva}

The software for the DP, DPM and alternative estimators of the Kullback-Leibler and the Hellinger divergence presented in this article are collected in a Python package which can be found in the repository at \url{https://github.com/statbiophys/catede}.
In addition, the package provides a Python version for the NSB entropy estimator \cite{nemenman_entropy_2001}, and a NSB estimator for the Simpson index \cite{simpson_measurement_1949}.

\newpage
\section{supplementary figures}
\label{sec:suppl-figures}
\setcounter{figure}{0}
\renewcommand{\thefigure}{S\arabic{figure}}
\begin{figure}
\begin{center}
\includegraphics[width=\columnwidth]{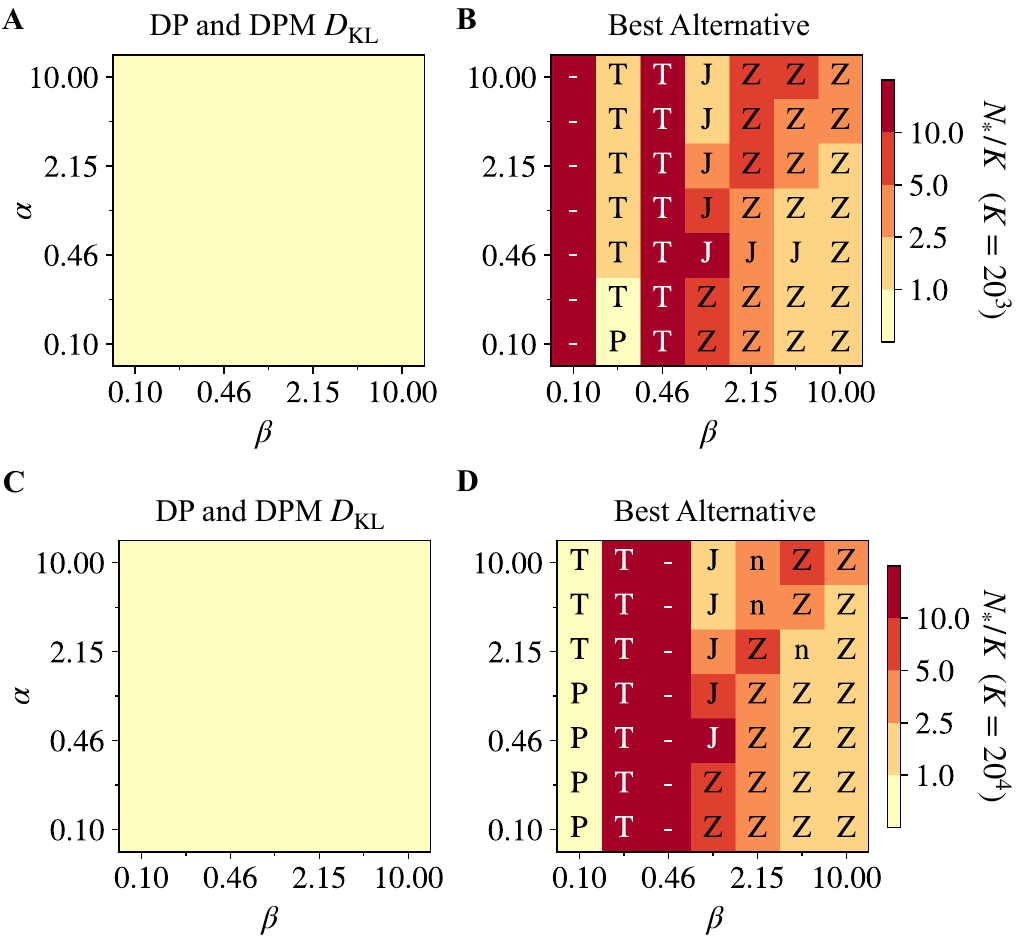}
\caption{
The convergence of the $\DKL$ estimates for different concentration parameters $\alpha,\beta$.
We use the same synthetic data of  Fig.~\ref{fig:dirichlet-convergence-estimation}B to plot the best score $N_{\ast}/K$ between DP and DPM for each combination of parameters.
We choose as a score $N_{\ast}/K$, where $N_{\ast}$ is the size $N=M$, at which the bias of the average estimate is smaller than $5\%$.
The average is computed over 30 repetitions.
\textbf{A.} 
Case $K=20^3$.
\textbf{B.}
The first letter of the name of best alternative method or a symbol ``-'' if no method converges for $N_{\ast}/K<50$.
The DP and DPM always outperforms the best alternative in the tested cases.
\textbf{C.} 
DP and DPM convergence of Fig.~\ref{fig-si:dkl-dirichlet-larger-K}A for the case $K=20^4$.
\textbf{D.} 
Analogous of Fig.~\ref{fig-si:dkl-dirichlet-larger-K}B for the case $K=20^4$.
}
\label{fig-si:dkl-dirichlet-larger-K}  
\end{center}
\end{figure}
 
\begin{figure}
\begin{center}
\includegraphics[width=\columnwidth]{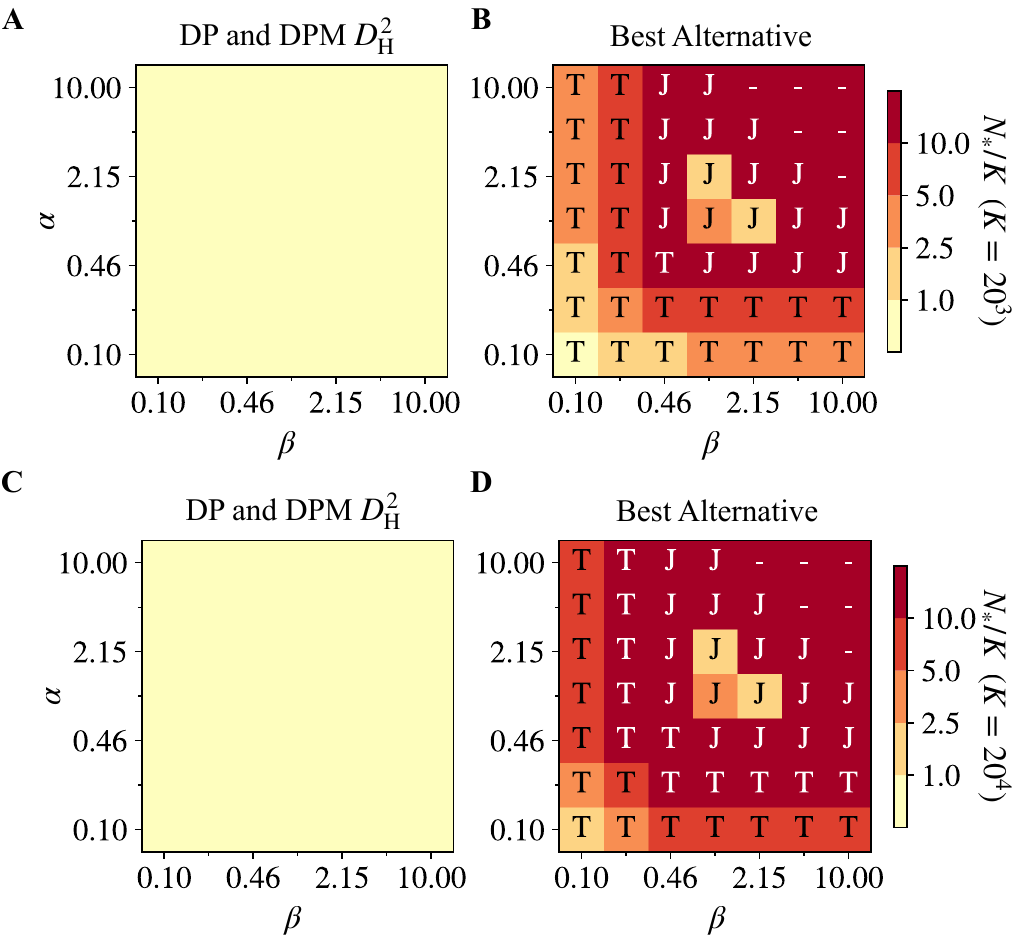}
\caption{
The convergence of the $\Hell^2$ estimates for different concentration parameters $\alpha$, $\beta$.
These figures use the same synthetic samples as in Fig.~\ref{fig:test-for-hellinger}A and correspond to the case $K=20^3$ and $K=20^4$.
Captions to \ref{fig-si:dh-dirichlet-larger-K}A, B, C and D are analogous to Fig.~\ref{fig-si:dkl-dirichlet-larger-K}.
}
\label{fig-si:dh-dirichlet-larger-K}  
\end{center}
\end{figure}

\end{document}